\documentclass[10pt,twocolumn]{article} 

\usepackage{wrapfig}

\usepackage{simpleConference}
\usepackage{times}
\usepackage{graphicx}
\usepackage{amssymb}
\usepackage{url,hyperref}
\usepackage{lipsum}
\usepackage{algorithmic}
\usepackage{caption}
\usepackage{array}
\usepackage{xcolor}
\usepackage{grffile}
\usepackage{textcomp}
\usepackage{stfloats}
\usepackage{url}
\usepackage{verbatim}
\usepackage{graphicx}

\begin{document}

\title{5G/6G-Enabled Metaverse Technologies: Taxonomy, Applications, and Open Security Challenges with Future Research Directions }

\author{Muhammad Adil$^{1}$, Houbing~Song$^{2}$, Muhammad Khurram Khan$^{3}$, Ahmed~Farouk$^{4}$, Zhanpeng Jin$^{5}$\\
\\
$^{1,2}$Department of Information Systems, University of Maryland, Baltimore County, Baltimore, MD 21250 USA \\
$^{3}$Center of Excellence in Information Assurance, King Saud University, 011 Riyadh, Saudi Arabia \\
$^{4}$ Computer Science and Physics Department, Wilfrid Laurier University, Canada \\
$^{5}$Department of Computer Science and Engineering, University at Buffalo,  NY 14260, USA\\
\\
(muhammad.adil@ieee.org, h.song@ieee.org, mkhurram@ksu.edu.sa, Afarouk@wlu.ca, zjin@buffalo.edu)  \\
\\
}

\maketitle
\thispagestyle{empty}

\begin{abstract}
\textbf{Internet technology has proven to be a vital contributor to many cutting-edge innovations that have given humans access to interact virtually with objects. Until now, numerous virtual systems had been developed for digital transformation to enable access to thousands of services and applications that range from virtual gaming to social networks. However, the majority of these systems lack to maintain consistency during interconnectivity and communication. To explore this discussion, in the recent past a new term, \textit{"Metaverse"} has been introduced, which is the combination of "meta" and "universe" that describes a shared virtual environment, where a number of technologies, such as fourth-generation ($4^{th}$) and fifth-generation ($5^{th}$) technologies, virtual reality, machine learning algorithms, artificial intelligence, etc, work collectively to support each other for the sake of one objective, which is the virtual accessibility of objects via one network platform.   With the development, integration, and virtualization of technologies, a lot of improvement in daily life applications is expected, but at the same time, there is a big challenge for the research community to secure this platform from external and external threats, because this technology is exposed to many cybersecurity attacks.   
Hence, it is imperative to systematically review and understand the taxonomy, applications, open security challenges, and future research directions of the emerging Metaverse technologies. In this paper, we have made useful efforts to present a comprehensive survey regarding Metaverse technology by taking into account the aforesaid parameters. 
Following this, in the initial phase, we will explore the future of  \textit{Metaverse} in the presence of $4^{th}$ and $5^{th}$ generation technologies in the context of interoperability and security. Thereafter, we will discuss the possible attacks to set a preface for the open security challenges. Based on underlined security challenges, we will suggest potential research directions that could be beneficial to address these challenges cost-effectively.  }
\end{abstract}

\section*{Keywords: } Metaverse technology, Security challenges, 5G/6G technologies, virtual reality, data privacy and preservation 

\section{Introduction}
In 1992, Neal Stephenson introduced the term ``metaverse'' for the first time in his book ``Snow Crash'' \cite{Kim2021}. He described the Metaverse as a vast virtual environment where objects are interlinked with the real world and people interact with them via digital avatars. Since its introduction, the Metaverse as a virtual universe has been described by a wide range of ideas that include life-logging technologies, spatial internet and embodied internet, collective virtual space, and a mirror world of simulation and collaboration \cite{Song2021, Owens2011, Balica2022}. Since Mark Zuckerberg, the Chief Executive Officer (CEO), affirmed Facebook’s rebranding as Meta in October 2021, the brilliant idea behind the new name has gained a lot of attention on social media and sparked a lot of discussions among diverse communities such as academia, enterprise market stakeholders, industry workers, and experts, etc. Besides Meta, several IT sectors have also shown strong interest in this business and are now investing in developing a Metaverse. To exemplify this, Microsoft recently purchased a video game holding company named Activision Blizzard as a piece of the contract to extend virtual gaming or online technology into the Metaverse.

Following the exponential growth of Metaverse, it is likely expected that in the near future, this technology would be pinpointed as a game-changer, because it getting the attention of major stakeholders such as Internet finance companies, online gaming companies, social networks, and many more leading technologies. 
To continue, the Seoul local government (South Korea) recently announced the Metaverse plan with the intention of creating a virtual communication paradigm for all municipal administrative areas, such as culture, economy, tourism, civic services, and educational activities \cite{Um2022}. Likewise, the report prepared by Bloomberg Intelligence in 2020 \cite{Visconti2022}, highlights that Metaverse revenue will increase from USD \$500 billion to USD \$800 billion in 2024,  of which half will be from the online game industry. Following this huge revenue, traditional video game companies are planning to shift from their existing gaming framework to a three-dimensional (3D) environment/virtual world in coordination with social networks \cite{Ko2021}. Furthermore, some additional new attractive activities, e.g. live entertainment, live gaming, media advertising, and other important events, could further enhance the utilization of this technology in the future \cite{Kshetri2022}.

At this point, we are considerably optimistic that the Metaverse technology will contribute significantly in many sectors in the future. However, on the other side, their intrinsic reliance on extensive connectivity and communications exposes them to several security threats that would be damaging factors for its interested stakeholders. Therefore, the security of Metaverse technology requires the special attention of concerned experts, network engineers, and research community stakeholders to gain the trust of clients and enterprise market users. To this end, some efforts have been made to address different security problems. However, the inherent factors, like heterogeneity, dynamic communication, unstructured deployment, interconnectivity of different devices, etc., still offer many security challenges that need to be addressed for the foolproof security of this technology.

To explore this area in the context of future applications, applicability, and expandability of Metaverse technology, we introduce in this paper a useful systematic survey regarding the security concerns of Metaverse applications while taking the future of this technology into account. Before moving on to the detailed discussion, we will first familiarize the readers with the taxonomy and different applications of Metaverse technology. Once the preface to understanding is set, we will extend our discussion to the security concerns of Metaverse technology by examining various internal and external threats. Consequently, we will move forward to highlight some open security challenges of this emerging technology with future research directions.

\begin{figure*}[ht!]
	\centering
	\includegraphics[width=.95\linewidth]{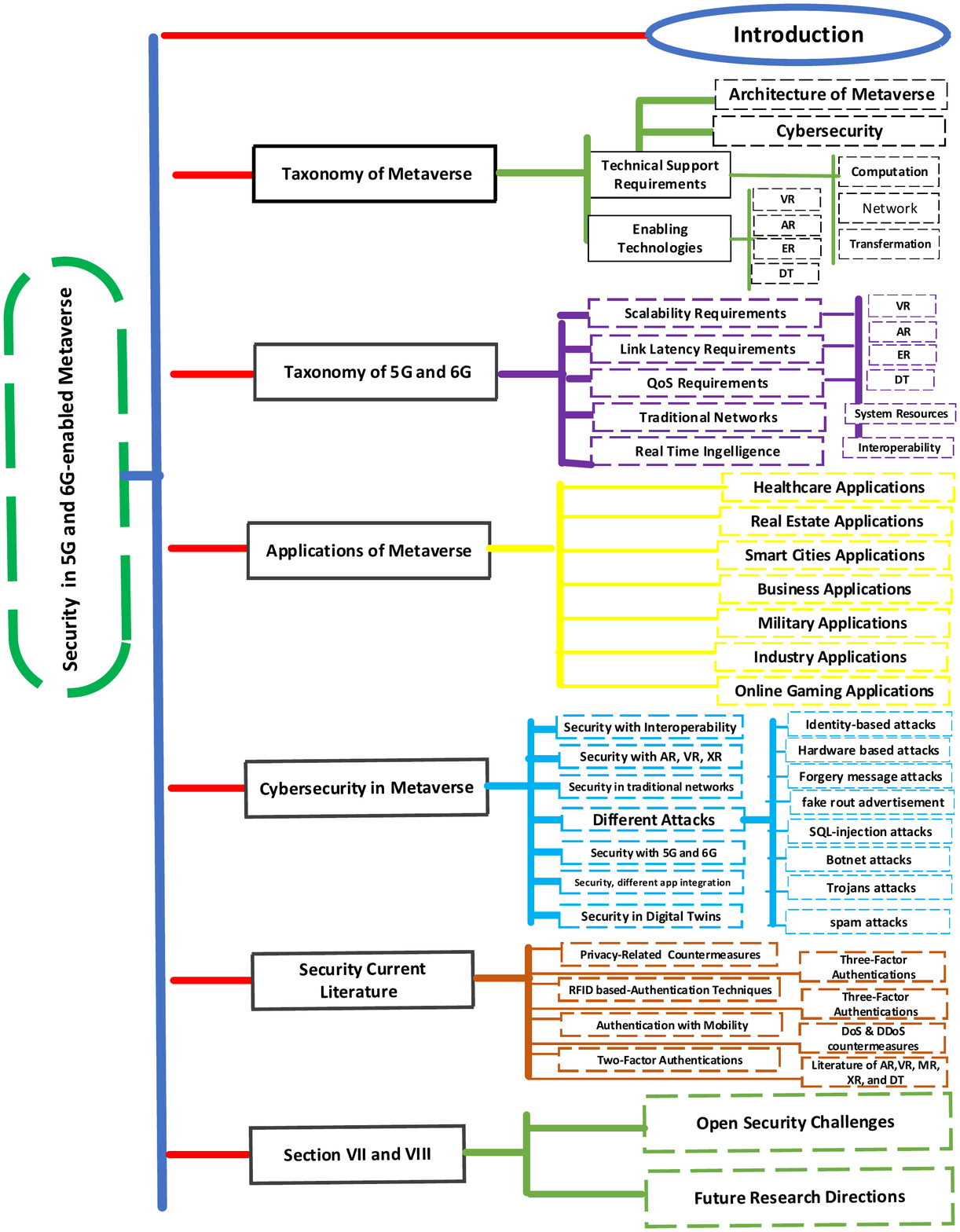}
	\caption{ Paper organization flowchart }
	\label{flowchart: 1}
\end{figure*}

The key contributions of this systematic review article are summarized below:

\begin{enumerate}
	\item Initially, we will begin our discussion with the taxonomy of the Metaverse technology and its current applications followed by future domains. Thereafter, we will talk about 5G/6G technology by taking into account the operational capabilities and requirements of Metaverse technology. 
	
	\item Following that, we will go over various review articles associated with Metaverse technology. To the best of our knowledge, there is not a single comprehensive survey paper in the literature that focuses on the security concerns of this emerging technology. We aim to ensure the novelty of this work by discussing the latest review articles to set the stage for the comparative analysis.
	
	\item  
	Next, we will discuss the possible security threats that can be exercised/applied to this emerging technology.
	In correlation with the highlighted threats, we will cover relevant literature that has been used to tackle these security attacks. But at the same time, we will highlight the limitations of different counteraction techniques by considering 5G/6G technology interconnectivity, interoperability, and communication aspects.  
	
	\item  In light of the limitations underscored in the literature and the requirements of Metaverse technology, we will talk about the open security challenges with future research insights. With this, we will give a concrete road map to the concerned stakeholders to enhance the productivity of this emerging technology in the future with foolproof security. 
\end{enumerate}
The rest of this survey paper is organized as below: Section I familiarizes the readers with the basic introduction of Metaverse technology followed by the key contributions of this work. In Section II, we discuss the taxonomy of Metaverse technology by taking into account different network and communication metrics. Likewise, Section III, of the paper contains the introduction of the 5G and 6G technologies paradigm, whereas Section IV overviews the application of Metaverse technology. In Sections V, VI, and VII, we discuss different security threats, current countermeasure schemes followed by the security requirements, and challenges of Metaverse technology.  The open security challenges are discussed in Section VIII, while the future research directions are outlined in Section IX. Section X summarizes and concludes the paper.  Flowchart of the overall paper organization \ref{flowchart: 1}.

Symbols and abbreviations used in this paper are summarize in table \ref{tab11}.

\begin{table*}[ht!]
	\caption{List of abbreviations and notations with their full description}
	\begin{center}
		\small
		\begin{tabular}{p{2cm}|p{6cm} |p{2cm}  |p{6cm} }
			\hline
			\cline{2-4} 
			\textbf{abbreviations} & \textbf{\textit{Full Description }}& \textbf{\textit{abbreviations}}& \textbf{\textit{Full Description }}   \\
			
			\hline CEO & Chief Executive Officer & IT & Information Technology \\
			
			\hline ISO & International Organization for Standardization & API & application programming interfaces \\
			
			\hline XR & Extended Reality & HCI & Human-Computer Interaction  \\
			
			\hline BCI & Brain-Computer Interface & QoE &  Quality of Everything \\
			
			\hline DT & Digital Twin & AI & Artificial Intelligence \\
			
			\hline GPU & Graphics Processing Unit & VR & Virtual Reality  \\
			
			\hline AR & Augmented Reality  & MR & Mixed Reality \\
			
			\hline PGC &  professional-generated
			content  & UGC & user-generated content  \\
			
			AIGC & artificial
			Intelligence-Generated Content  & NFT & non-fungible token \\
			
			\hline mMTC & Massive Machine-Type Communication  &  eMBB & Enhanced Mobile Broadband \\

			SDNs & Software-Defined Networks & DoS & Denial of Service \\
			
			\hline DDoS & distributed denial of service & IoT & Internet of Things  \\
			
			\hline  CP & cyclic prefix & WSN & Wireless Sensor Network \\
			
			\hline MRI & Magnetic Resonance Imaging & CT & Computed Tomography \\ 
			
			\hline IoV & Internet of Vehicles & TAR & Technology-assisted Review \\
			
			\hline GPS &  Global Positioning System & STE & Synthetic Training Environment  \\
			
			NFV & Network Functions Virtualization & MAC & Media Access Control  \\
			
			\hline ML & Machine Learning & URLLC & Ultra-Reliable and Low Latency Communication\\
			
			\hline UAVs & Unmanned Aerial Vehicle  & KVI & Key-Value Indicators  \\
			
			\hline KPIs & Key Performance Indicators & UN & United Nation \\
			
			\hline SDG &  Sustainable Development Goals & TEE & Trusted Execution Environments  \\
			
			\hline SIM & subscriber
			identity module  &  SCM & System-on-Chip Module \\
			
			\hline  PERT & Privacy Enhanced Retrieval Technology &  PPDC & Privacy Preserving Data Collection \\
			
			\hline  RFID & Radio Frequency Identification & PUF & Physically Unclonable Functions \\
			
			\hline AES & Encryption Standards & ECDA &  Elliptic Curve Digital Signature Algorithm \\
			
			\hline  GSM & Mobile Communication Model & ECC & Elliptic Curve Cryptography \\
			
			\hline AKE & Authenticated Key Exchange & AVISPA & Automated Validation of Internet
			Security Protocols and Applications \\
			\hline  BAN & Burrows-Abadi-Needham &  NR-PMT & non-Repudiation Private Membership Test \\
			
			\hline CRN & Cognitive Radio Networks & ZKP & Zero-Knowledge proof \\
			
			\hline GAN & Generative Adversarial Network & ATM & Automated Teller Machine  \\
			
			\hline NCC & Network Control Center & CPS & Cyber-Physical Systems \\ 
			
			\hline TLS & Transport Layer Security & AHM & Authentication Handover Module \\
			
			\hline IoE &  Internet of Everything & VPPM & Virtual-Physical Perceptual Manipulations \\
			
			\hline IoMT & Internet of Medical Things   & KD & Key Distribution \\

			\hline
			
		\end{tabular}
		\label{tab11}
	\end{center}
\end{table*}

\section{Taxonomy of Metaverse }

\begin{figure*}[ht!]
	\centering
	\includegraphics[width=.95\linewidth, height = 15 cm]{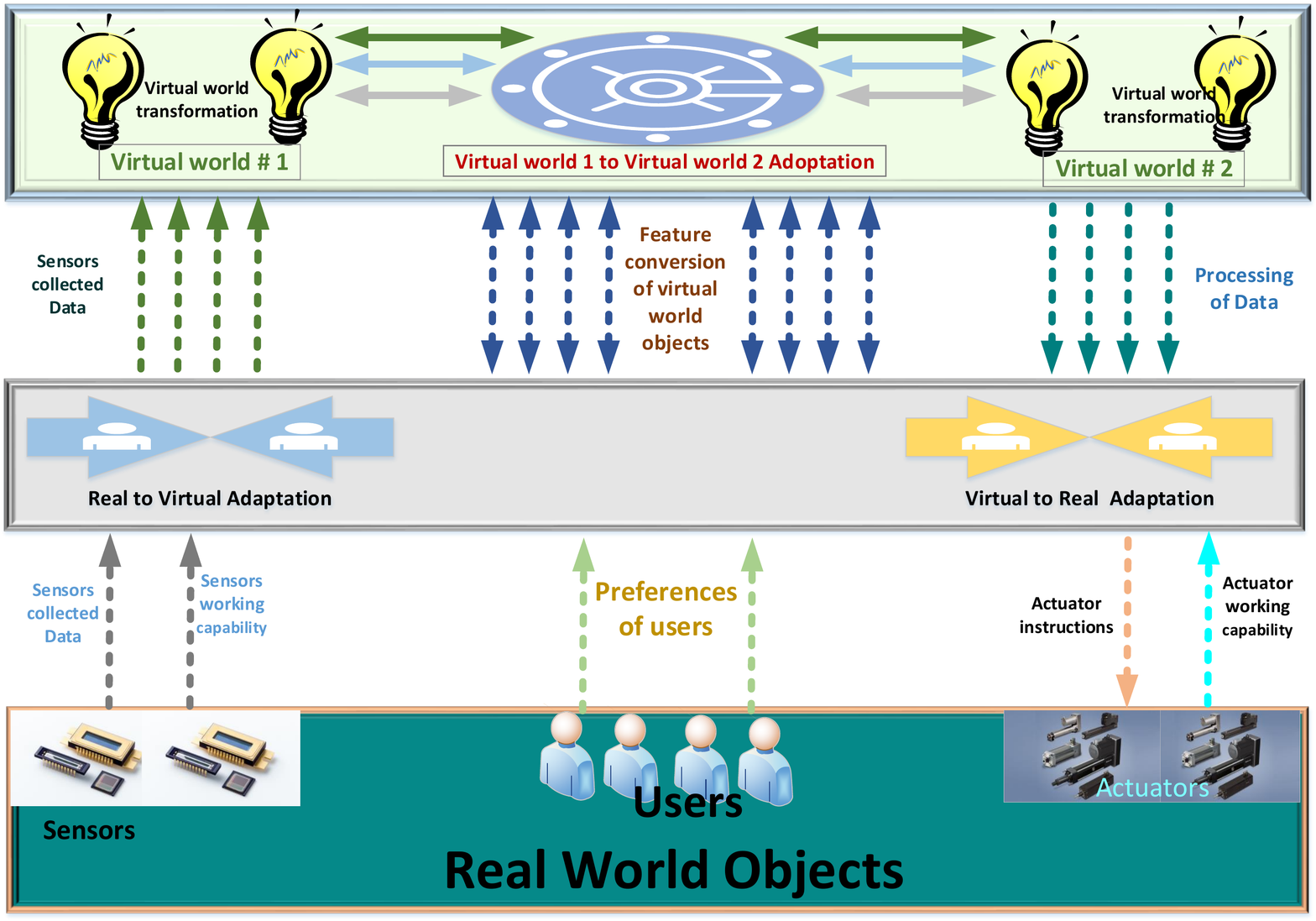}
	\caption{ Operational steps in Metaverse technology paradigm }
	\label{fig: 1}
\end{figure*}

In this section, we will familiarize the readers with Metaverse technology by going through its existing standards, network architecture, fundamental features, potential applications, and enabling technologies. To continue this discussion, ISO/IEC 23005 (MPEGV)was used as the first standardization prototype in the Metaverse technology for networked virtual environments (NVEs) to create an interface between the virtual world objects and real-world objects for the sake of interconnectivity and seamless communication \cite{ISO/IEC23005}. In 2011, the first version of this prototype was released, while its $4^{th}$ edition was published in 2020. Likewise, the suggested prototype is appropriate for several business applications and services interlinked with the Metaverse technology. It also handles sensory effects, audiovisual information, and characteristics of the virtual objects to satisfy the interaction needs between virtual objects and real-world objects. To generalize the operational and media exchange scenario of this paradigm, we have used Figure \ref{fig: 1} for illustration. 

To explore Figure \ref{fig: 1}, the sensors, intelligent appliances, and Internet-of-Things devices collect information from the real world and process it according to the internal processes of ISO/IEC 23005 (MPEGV) to convert it to virtual world objects. Thereafter, these processes move one step forward and convert the media from the virtual world to the real world by considering the input commands of actuators followed by the preferences of users. Finally, continuous data or media adoption is established between the real world to the virtual world, the virtual world to the virtual world, and the virtual world to the real world.         
To summarize, the ISO/IEC 23005 standards primarily concentrate on the sensory data effects, which demonstrate that these standard lacks to offer general-purpose interfaces between the virtual world and real-world items, objects, and entities, etc.

In 2019, the IEEE 2888 project was launched to define standardized interfaces for the synchronization of cyber and physical worlds \cite{IEEE2888}. This standard provides a foundation for Metaverse technology interconnectivity by setting information formats and APIs (application programming interfaces) for controlling actuators to efficiently collect and process sensory data in the network. Likewise, the vendor released an updated version of this paradigm with time to improve the operation and productivity of Metaverse technology.

\subsection{Architecture of Metaverse Technology}

\begin{figure*}[ht!]
	\centering
	\includegraphics[width=.98\linewidth, height = 16.5 cm]{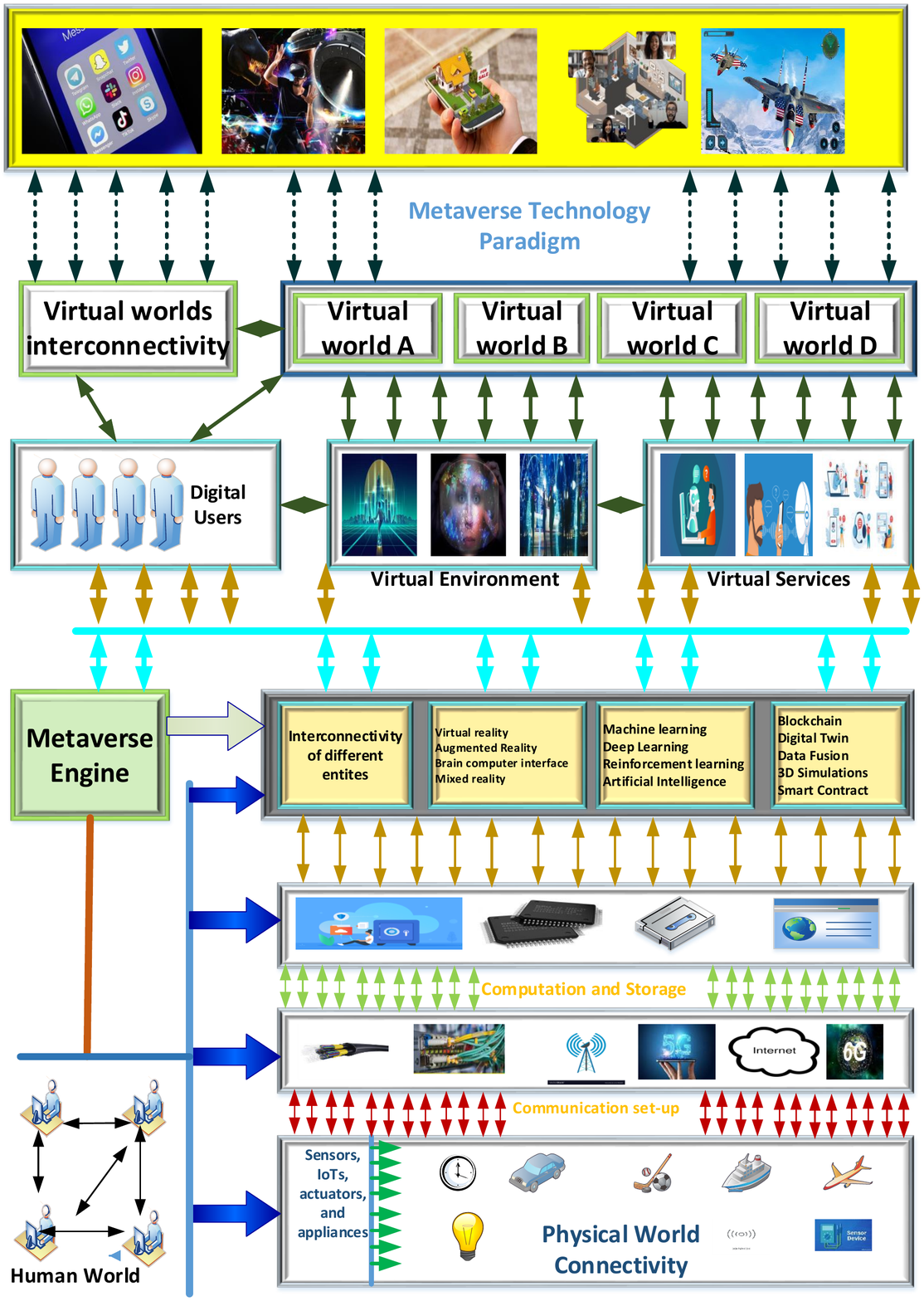}
	\caption{ Metaverse technology paradigm }
	\label{fig : 2}
\end{figure*}

Metaverse is an emerging technology paradigm that has self-sustaining and hyper-configurational capabilities with a 3D virtual transferred environment, which is formed by the confluence of virtually enhanced physical reality and physically persistent virtual space \cite{Mystakidis2022}. To generalize, Metaverse is a synthetic world that is made up of the user's controlled actuators, avatars, sensors, digital objects, virtual objects, and computer-generated elements, etc., where people/avatars interact, play, collaborate, watch, and socialize with each other utilizing smart devices.
Likewise, the collection of physical objects, humans, and digital ternary realms leads to the creation of a Metaverse technology paradigm. In summary, we demonstrate a visual representation of the Metaverse technology paradigm in Figure \ref{fig : 2}.

From Figure \ref{fig : 2}, it is clearly understood that Metaverse technology is human-centric, whereas human beings in coordination with social and psychological interactions constitute a real-to-virtual world. This virtual world is operated with the help of smart sensors, wearable devices, actuators, avatars, and virtual entities through extended reality (XR) technologies and human-computer interaction (HCI). On the other side, the physical world of the Metaverse technology paradigm creates an opportunity to establish sensing, controlling, perception, processing, and transmission between interconnected sensory devices of the physical world \cite{Scavarelli2021}. With this, an effective interaction infrastructure is developed between the digital world and the human world. 

Following this, IEC 23005 and IEEE 2888 standards interconnect a number of devices to form real object networks that lead to distributed virtual worlds (sub-metaverses) \cite{Kraus2022}. These sub-metaverses (virtual worlds) can nourish users' illustrated digital avatars with certain classes of virtual goods and services such as social dating, online gaming, online museums, markets, and online concerts, etc.
Based on the aforementioned arguments, the main sources of information in the Metaverse technology paradigm include: 1) the collection of information from the real world (input) and 2) the transformation of real-world information to the virtual world (output).

\subsection{ Technical Support Requirements of Metaverse }

From the architecture of Metaverse, we can see that this technology is constituted from a layer-wise paradigm, which includes the interaction layer, supporting technologies, network layer, and application layer \cite{Wang2021, Zhang2022}. To explore the key objective of the interaction layer is to establish communication between humans, physical world, and virtual world technologies such as robots, brain-computer interface (BCI), and XR etc., with an excellent quality of everything (QoE) users experience \cite{Xu2022}. 
Likewise, the network layer handles wire and wireless data transmission among network interconnected entities. Problems in the network layer affect the creation, innovation, and execution of products and services in the Metaverse because they exist in the form of latency, transmission delay, security, and throughput \cite{Yang2022}. Following this, we can see from the literature that the majority of emerging technologies use 4G technologies, as a main network paradigm throughout the globe, and this could be for Metaverse as well. Likewise, the application layer of Metaverse is dealing with content production, execution, and maintenance followed by spatial mapping, session initiation, generation, and authentication \cite{Wang2022}. Given that, spatial mapping will ensure the real-world objects mapping related to technologies such as digital twin (DT) and holography in real time. Similarly, content generation has great importance in Metaverse technology, and is assumed the driving force behind the operation of this technology, because it deals with Artificial Intelligence (AI) and visualization of technologies \cite{Christodoulou2022}. Next, in Metaverse the authentication of devices, users, and messages is ensured through standard protocols and underlying technologies for the realization, but still, the research is in progress to guarantee the foolproof security of this technology.

\subsubsection{Network Requirements}

In the future, we are pretty sure that communication technologies for Metaverse would be facilitated through 5G and 6G communication paradigm. Therefore, it is imperative for researchers and industry stakeholders to build network entities and software that could be capable to support this ultra-fast communication platform. Furthermore, Metaverse could be a super-large virtual twin forum, where the users will expect delay-sensitive communication for their social, personal, and business need \cite{Fu2022}. At present, such a high demand-oriented delay-sensitive application does not exist, but in the future Metaverse, there will be, because social, personal, and technical activities of the sub-metaverse will be sensitive to minor errors, synchronization issues and intruder interruption \cite{Mystakidis2022}. Despite that, the realization of interoperability aspects among tactile synergy, XR, holographic perception, and communication followed by other technologies. Indeed, latency, throughput, and security are the hardest parameters of the Metaverse network that need proper redressal techniques in the design and execution phase.

\subsubsection{Computational Requirements}

As discussed in the network requirements, future 5G and 6G-enabled Metaverse would demand delay-sensitive communication. For the satisfaction of this, one challenging problem should be the low computation capabilities of existing hardware and software, because this technology will require more powerful computation entities to fulfill the needs of network and users in the context of data transmission, reception, and sub-metaverses synchronization \cite{Lee2021}. With this, the construction of large-scale complex digital twin and digital human space initiates some basic problems such as system power, heat, and fluids, etc., that require the communal efforts of many stakeholders. Given that, these actions should be used in a way to assure computation friendly paradigm of the Metaverse by using numerical calculations, simulations, and visualizations use cases.

Therefore, we believe that the development of computation-friendly hardware followed by edge and cloud computing infrastructure can improve intra-system and inter-system latency with better power consumption statistics. To exemplify, GPU computational capability may improve the visual impact of Metaverse and cloud games by allowing the concerned market stakeholders for more realistic settings and things to be modeled \cite{Alam2022}. And edge and cloud computing may be useful to enhance the performance of terminal equipment such as networked devices and data centers because they will not transfer data back to the central server. As a result, the overall processing speed increased with the least response time of users.

\subsubsection{Real World and Virtual World interactive interface}

In real life, the Metaverse should be easily accessible to users, operators, and administrators for operations. Presently, this technology uses virtual reality (VR), extended reality (XR), mixed reality (MR), augmented reality (AR), robots, systems, digital twins, and BCI to connect realities (real and virtual) under one network platform  \cite{Han2022}. To explore XR, it is the advanced form of VR, AR, and MR. Similarly, VR is exploited as an emerging technology to introduce users to a virtual world/environment that is identical or different from the real world. While AR imposes and supervises a layer of virtual information based on the preservation of actual world information \cite{Burova2022}. This allows users to obtain data and information that had been assessed by computers in real-time, and could be useful in their job and decision-making process. However, MR refers to a visual environment that is formed by the integration of the real world and the virtual world followed by VR and AR. XR technology as a whole recognizes the input information of users by producing a specific or desired output. In general, XR technologies work as a supportive platform to ensure the interaction of the virtual and real world, where the users may get a chance to view objects in digital format \cite{Zhang2022}. At present, most of XR technologies relies on traditional control systems such as touch and smell, which is going through many security problems. To compensate these problems, and establish long term stable virtual world, some new research areas are needed to be explored.


\subsection{Content Creation and Execution}
	
Content generation, execution, presentation, and assessment is an important parts of Metaverse \cite{Tlili2022}. To develop, a reliable and consistent Metaverse, a lot of simulation and computation are needed to make the real world and virtual world compatible to each other.  Given that, synchronizing and simulating the real world with the virtual world is the first step in constructing the Metaverse, where the users will share a massive amount of data with the help of content. Although this technology is newborn, but, it has shown some valuable results in the recent past with AI. Content generation and execution ranges from general to specific such as professional-generated content (PGC), user-generated content (UGC), and artificial intelligence-generated content (AIGC) \cite{Duan2022}. Even though the cost of content design and innovation is decreasing, but, the efforts of technology requirements and intelligence are gradually increasing. Following this, the content generated by AI is more important and productive, because it has a lot of potential to enhance the productivity of future industries \cite{Huang2022}. But at the same time, it is susceptible to complex attacks, which require new security hardware and software-based techniques that could be compatible with content creation, execution, and presentation.

\subsection{Cybersecurity}

Metaverse demands a completely digital platform, which requires a special security platform, unlike traditional human societies to ensure the security of different consumers. Therefore, the core objective of this paper is to emphasize on the security aspects of this technology. Herein, we superficially evaluated the security requirements of Metaverse \cite{Wang2022}. At present, conventional applications use centralized authentication schemes with hard codes, passwords, and identity frameworks, which could not be useful for large-scale Metaverse. Keeping in view the diversity, heterogeneity, and virtuality of the Metaverse, a specific credit system should be developed for this technology. We explored this topic from Section V onward in this paper.

\subsection{Superficial Evaluation of Enabling Technologies of Metaverse Paradigm}

In this subsection, we will discuss the enabling technologies associated with the Metaverse technology paradigm. With the help of these technologies, the Internet-connected ecosystem in terms of real-world and virtual-world objects is navigated and controlled by avatars/users. However, mapping such a large number of entities requires an intelligent framework to ensure reliable results. To ensure reliable operation, the Metaverse technology paradigm uses a range of services endorsed by many interconnected cutting-edge technologies. These technologies are superficially summarized in Table \ref{tab1}, and comprehensively discussed in the upcoming sub-sections of this sections.

\begin{table*}[htbp]
	\caption{Edge-cutting enabling technologies of the Metaverse paradigm}
	\begin{center}
		\small
		\begin{tabular}{p{1.8cm}|p{5.5cm} |p{5.3cm}  |p{4.8cm} }
			\hline
			\cline{2-4} 
			\textbf{Name of technology} & \textbf{\textit{Description and Relevancy}}& \textbf{\textit{Advantages}}& \textbf{\textit{Concerned problems}}   \\
			
			\hline Blockchain Technology, references \cite{Lv2022, Gadekallu2022} & Metaverse technology paradigm enables the trading of virtual assets that involves interaction with the real economy. Therefore, managing and securing this technology in a decentralized and heterogeneous environment demands steady and transparent trading mechanisms to ensure reliable operation. To deal with this, blockchain technology should be used as an effective paradigm.    
			& With the utilization of blockchain technology in the Metaverse paradigm, many centralized trading problems can be addressed. Secondly, it is the most promising technology to manage non-fungible token (NFT) trading problems in a decentralized environment. Moreover, it also facilitates peer-to-peer trading of digital assets in a distributed and decentralized environment. & In contrast, this blockchain technology offers several security challenges when it comes to ultra-fast communication such as 5G/6G. Moreover, decentralized security management raises computational complexity problems on the edge side.  \\
			
			\hline Artificial Intelligence \cite{Gadekallu2022, zvarikova2022retail}  & Artificial intelligence had shown        incredible results in emerging technologies, and it can be a game-changer in the Metaverse technology paradigm because it will empower them in many aspects. This technology performs really well, when it learns from the data experience, and the same is true with Metaverse technology.  
			& To generalize the importance of artificial intelligence in the Metaverse technology paradigm, it can enhance the key corporation processes with high-speed processing, accurate decision-making, secure data evaluation, etc. 
			& Conversely, secure firmware updates in a decentralized or heterogeneous network environment is a challenging task, because unlike vendors entities have different firmware update requirements. As a result, AI techniques exercise several problems.    \\
			
			\hline Machine Learning and Deep Learning \cite{kliestik2022live, xu2021wireless} & 
			
			In Metaverse technology, the role of Machine Learning, Deep Learning algorithms, and Reinforcement Learning can not be ignored, because these algorithms are proven and effective while tackling complicated tasks in different networks.  &  Machine learning, deep learning, and reinforcement learning play a paramount role in creating naturalistic virtual 3D worlds and automating them with a range of services in the Metaverse paradigm. To exemplify,  NVIDIA introduced GANverse3D which allows content creators to automatically generate virtual replicas of real-world objects photos. & In the Metaverse technology paradigm, the Training of models with ML, DL, and RL-enabled techniques is a challenging task because most of the operations, assessments, and decision-making processes are based on these models.   \\
			
			\hline Virtual Visualization Technologies \cite{gollob2022sensable, hudson2022virtual} & 
			Virtual visualization of technologies basically refers to virtual reality (VR), augmented reality (AR), and mixed reality (MR). To exemplify, these technologies had been used in head-mounted helmets, eyeglasses, and main display points are a demonstration of Metaverse technology. & This technology allows the users and clients to enjoy real-time interactive environments with virtual large-scale 3D modeling. The employed sensor, actuators, and IoT devices create a visually realistic virtual environment by sensing and sharing the surrounding object information in the network. & Although these technologies play incredibly well in the Metaverse paradigm, their interoperability raises very complex issues that hamper the expandability of Metaverse technology.  \\
			
			\hline Digital Twin \cite{adams2022virtual, zaman2022meet} & In the Metaverse technology paradigm, the role of the digital twin can not be neglected, because it represents the digital clone of objects, entities, humans, and systems in the real world with high consciousness.  & This technology usually encourages the mirroring of physical objects and entities to predict their virtual bodies by assessing the real-time data streaming of sensors IoT devices, users, actuators and physical models, etc. & To ensure reliable operation, self-learning, transfer learning, and self-adaption are the most challenging tasks for the research community to be managed. \\
			
			\hline Ubiquitous Computing \cite{zhang2022artificial, hudsonsmith2022ubiquitous} & 
			
			Ubiquitous computing provides consumers access to Metaverse technologies everywhere and at every time. It provides seamless interactions between human users and the physical space through ubiquitous smart gadgets implanted in the environment. & Metaverse technology paradigm, ubiquitous computing allows the human users/client to freely interact with the avatars and enjoy real-time Metaverse services via smart networked objects instead of personal computers or laptops. & Despite these advantages, it offers several security problems, because most of the time the end-user is not familiar with the security protocols of different applications. Therefore, they frequently do mistakes to compromise their security.  \\
			\hline Networking Technologies \cite{dhelim2022edge, tang2022roadmap} & In the Metaverse, the role of networking technologies such as 5G/6G, APIs, software-defined networks (SDN), and IoT devices can not be ignored, because they enable real-time and ultra-reliable communications between real and virtual worlds objects.  & With the integration of these technologies, a real and virtual world infrastructure should be established, where the human can play games, place different orders, improves their business, and many more.   & At the edge side of Metaverse technologies, 5G/6G-enabled ultra-fast communication is a challenging issue to be managed with proper security protocols. \\

			\hline
		\end{tabular}
		\label{tab1}
	\end{center}
\end{table*}

\subsubsection{Augmented Reality }

The term augmented reality (AR) originated in the late 1950s, but its applications had been seen in the $21^{st}$ century in the consumer market \cite{birlo2022utility}.  Generally, this technology works between the real world and the virtual world to differentiate between objects. Despite that AR work as a mobile technology, where users navigate between real space and virtual space. Given that AR ensures the transformation of information in the digital world by allowing users to learn and share about the real and virtual world statically, dynamically, and on-demand, etc. \cite{sun2022influence}. Keeping in view these factors, AR is considered the core technology in the development of Metaverse. From the textual discussion, it seems a very simple process, but in reality, it is very complex, and creates several problems that range from general communication aspects to particular network integration and security concerns. To explore, in Metaverse AR is assumed as a virtual content creation and execution paradigm that establishes the connection between the viewer, real world, and virtual world \cite{wang2022survey}. But these content intermediation process follows a variety of ways to enhance the real world and virtual world experience among users, tools and context, etc. Sometimes this intermediation link may be strong or weak, due to the user's location, and application environment in Metaverse. But occasionally, this link may be disturbed due to network susceptibility and illegal user interruption. Therefore, in this paper, we focused on this important issue to familiarize the readers, students, researchers, and involved stakeholders with current literature followed by future possible security threats and their counteraction directions. 

\subsubsection{Virtual Reality}

In the mid of 1980s, the term virtual reality (VR) was used by Jaron Lanier, when he started working on goggles and gloves by keeping in view the user's experience of the real world [45]. The basic notation behind this was to develop tools and testers for today's 3D world evaluation. 
In simple words, VR is the creation of a 3D world with the help of a computer that could be similar to the real world. To interact with this world, the users will use special devices such as goggles, gloves, clothes, and shoes, etc., to boost reality and immersiveness \cite{ott2023digital}. Given that, the movements of these devices and users will be detected via sensors simultaneously to ensure the proper illustration of the 3D world. 
Following the operational capabilities of VR, it is pretty clear that this technology has great weightage in Metaverse. Because it creates 3D spaces for us to participate in Metaverse, while AR plays the role of a bridge to confirm the connectivity of the virtual and real world. Therefore, both these technologies had great contributions to the Metaverse, and are expected more in the future.   Despite its numerous assistance, VR will offer some additional security threats that would be beyond the scope of existing literature. Therefore, we would like to explore these challenges in presence of the current literature to set the stage for the foolproof security of Metaverse.

\subsubsection{Mixed Reality}

Mixed Reality (MR) in general represents a realistic augmentation of the real world in the form of a virtual world \cite{singh2022augmented}. In this illustration, the users are unable to distinguish between virtual content and real objects. Although this technology is still in its emerging phase, and requires special hardware such as smart glasses, lenses, and screens to interact, track or view the environment \cite{eng2022new}. Despite that, this technology closely works with AR and VR to contribute to Metaverse, but still, it is in the progressive phase, and suffers from several interconnectivity, compatibility, and security issues. The main focus of this work is on security, therefore, we want to link and evaluated MR in the context of Metaverse. For this, we checked the current literature, and followed the limitations to highlight the open challenges and future research directions.

\subsubsection{Extended Reality}

Extended reality (XR) is used  to represent AR, VR, and MR collectively.  In the Metaverse, these technologies support each other to design and illustrate the real world and virtual world under one umbrella, where the differentiation between 3D objects and physical objects would be hard for users \cite{panda2022alltogether}. From the definition and contributions of XR, we can see that this technology has a key position in Metaverse. However, it is still challenging for the involved stakeholders to take care of the security of this technology, because it is in the progressive phase. Despite that, when it comes to the collective operation of AR, VR, MR, and other emerging technologies in Metaverse, the security problems get more severe. Therefore, we want to check these problems in the context of 5G and 6G communication paradigms by following the future of Metaverse. 

\subsubsection{Digital Twin}

In the previous subsections of enabling technologies, we noted that Metaverse will include digital transformation technologies followed by the visualization of integrated real-world and virtual worlds illustrations. However, the accurate resemblance of the real world to the virtual world is a big problem. To tackle this issue, Digital twin (DT) has emerged as an enabling technology that helps to represent physical objects in the digital mirror \cite{cai2022compute}. Likewise, this technology ensures the synchronous evolution of twins throughout the entire process, and it is difficult for spectators to differentiate between physical and virtual objects \cite{matthie2022use}. Despite the fact that this technology has significant assistance in the Metaverse, and could be useful in the future as well. But once, it comes to the adaptation of DT in Metaverse, then, there are several challenges, which need sufficient attention from the concerned authorities in the redressal process. Given that, every challenge has its own consequences, but here in this paper, we will focus on the security aspects of DT by keeping in view the Metaverse.

\subsection{Summary of Discussion}

To summarize, in this section, we talked about the architecture followed by the fundamental requirements of Metaverse. The basic notation behind this segment was to familiarize the readers, students, and researchers with Metaverse technology. Given that, we assumed the circumstances of general people interested in this technology rather than experts. Therefore, we give a superficial touch to each aspect that can be helpful for the new learners of this technology. Moreover, we are interested to set a foundation for the security concerns of this technology, which is the basic objective of this work. Therefore, it was important to give an overview of the taxonomy of Metaverse followed by its enabling technologies

\section{5G and 6G Technologies Taxonomy}

In this section, we will look at different aspects of 5G and 6G technology and how well this technology will perform in the global markets recently by particularly taking into account the Metaverse paradigm. With this, we will set a preface for the 5G/6G-enabled Metaverse technology paradigm. But before diving into this discussion, it is pertinent to evaluate the key components interlinked with 5G and 6G technology. 5G and 6G-enabled technologies use small cells to manage network densification problems and enhance the coverage area with a high transmission rate \cite{mahdi2021from}. To handle ultra-fast communication with these technology-enabled environments or applications, carrier aggregation should be used as an alternative network component, which has the capability to ensure the facilitation of clients/end-users with more than a single-component carrier to offer higher bandwidth services \cite{qadir2022towards}. 
Despite the advantages of carrier aggregation, it has some implications, because it supports several frequency bands at a time, which can impact the end-user in terms of connectivity and services. For redressal of this problem, a radio access network (C-RAN) should be used as a substitute technology for the 5G and 6G-enabled Metaverse networks because it has the ability to get rid of hardware issues associated with the end-user connectivity and services \cite{chowdhury2019role}. However, this technology also creates scalability concerns, when large-scale network expandability is expected because it raises computation and calculation issues on the client side.  
When it comes to the client-side computations and calculations, the role of software-defined networks (SDNs) and SDN controllers can not be ignored in 5G and 6G-enabled Metaverse technologies, because they are computation-friendly on the client side \cite{azari2021evolution}. In contrast, SDN is less effective in managing the security concerns in heterogeneous networks such as $\textit{Metaverse}$, because they are unable to establish a trusty relation among the user's access applications and SDN controllers.

\begin{figure*}[ht!]
	\centering
	\includegraphics[width=.95\linewidth, height = 16 cm]{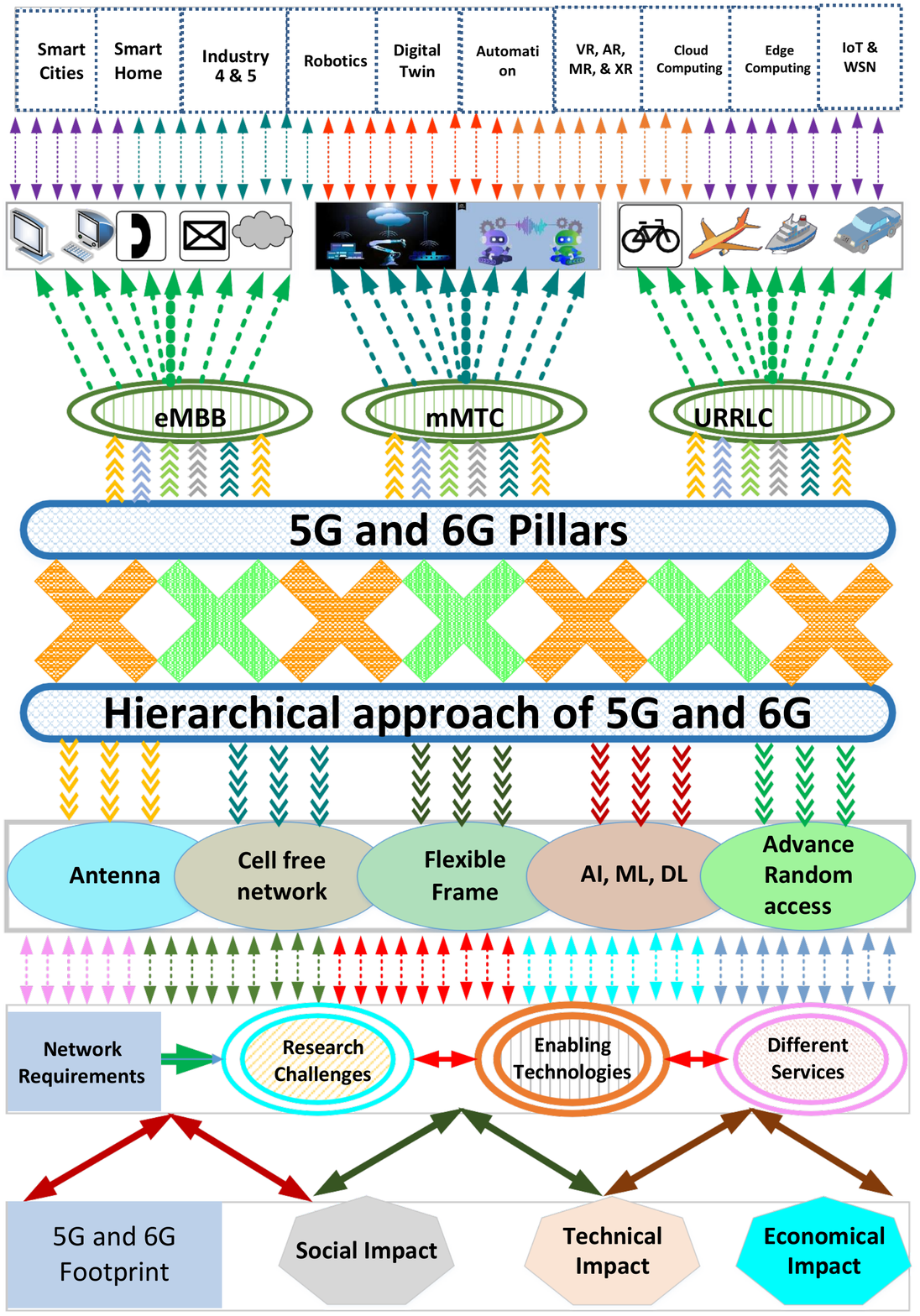}
	\caption{ Hierarchical approaches with 5G and 6G network paradigms for Metaverse technologies applications }
	\label{fig : 3}
\end{figure*}

To explore the security issues in the 5G and 6G-enabled Metaverse technology paradigm, it has been noted that these networks are not immune to distributed denial of service (DDoS) attacks, eavesdropping attacks, message forgery attacks, device tampering attacks, and many more \cite{gai2021summary}. Therefore, it is important to improve the security standards, policies, and protocols to ensure the security of billions and millions of interconnected actuators, cybers, sensors, IoTs, users, and other relevant entities in the network. In the consequent sections, we will shed light on the expected demands of 5G and 6G technology to set a background for the Metaverse technology interconnectivity. The hierarchical approaches with 5G and 6G network paradigms for Metaverse technologies applications are visualized in Figure \ref{fig : 3}.

\subsection{Communication Speed and Scalability in 5G/6G-enabled Metaverse Technologies}

In the recent future, it has been predicted that in the Metaverse technology paradigm machine-to-sensor, sensor-to-sensor, machine-to-machine, sensor-to-cyber, etc, connectivity in 2030 would be 670 times more than of the year of 2020 \cite{al-sarawi2020internet}. This development of this emerging technology inspires the academic community to think about the technical requirements of a variety of network problems that range from spectrum to communication and energy efficiency. In \cite{sabuj2022delay}, it has been reported that the 5G and 5G technology will facilitate massive machine-type communication (mMTC) with an enhanced mobile broadband (eMBB) connection up to 20 Gbps, as shown in Figure \ref{fig : 3}.

With the expected exponential growth of the Metaverse technologies, the anticipated limit of 5G would be reached by 2030. Therefore, the development of 6G technology is the utmost need of Metaverse technology. Accordingly, it is expected that the data transmission rate will be significantly improved with the induction of this technology up to 1 Tbps (up to 10 Tbps). However, 5G and 6G technologies use millimeter-wave transmission that ranges from 20–100 GHz, which causes several communication problems such as modulation, phase noise, non-linear power amplifiers, and analog-to-digital converter hardware and software, because of their high transmission and communication speeds. Therefore, we believe that in 5G/6G-enabled communication the frequency band would be beyond 100 GHz and perhaps would be up to a few THz \cite{saad2019vision}. To manage the communication processes in the future, the high data rates would be defined in the Metaverse technology by taking into account different services that are expected to be demanded by the clients in the near future. These services include AR, MR, human nano-chip implants, VR, autonomous systems, connected robotics, smart market, online gaming, online ordering, etc \cite{lee2021all}.  Despite some limitations of 5G and 6G technologies, it is believed that these technologies could be the best candidates to enable ultra-fast and reliable communication among interconnected millions and billions of devices in a Metaverse technology paradigm.  

\subsection{Link Latency in 5G/6G-enabled Metaverse Technologies}

In the Metaverse technology paradigm, several real-time services appear to be integrated as a part of this technology in the future. These services range from general applications to specific application development and deployment such as smart cities, smart homes, smart factories, autonomous vehicles, robotics, online gaming, etc, where human-to-cyber or physical world/virtual world interaction should be made with the help of new ways such as VR, AR, MR, and prosthetic limbs, etc. \cite{sheth2020taxonomy}.

To explore, most of the real-time services in Metaverse technology are time-sensitive that have defined latency requirements (10 milliseconds or below) to guarantee reliable communications among the interconnected sensor, cybers, humans, and other networking entities, etc. Furthermore, the networking entities and services associated with Metaverse technology can result in delay problems due to the length of the cyclic prefix (CP) in dedicated communication channels that require constant scheduling of traffic \cite{masaracchia2021uav}. To ensure delay-sensitive communications in the Metaverse paradigm, 5G and 6G technologies should be used as the best candidates for this purpose. However, the link establishment and maintenance among interconnected devices will cause several problems with such ultra-fast communication. Therefore, we need to reevaluate the existing communication equipment followed by the physical, network, and application layer protocols that are expected to be used in this technology.    

\subsection{Quality of Service in 5G/6G-enabled Metaverse Technologies }

In the Metaverse, lightweight devices such as head-mounted displays (HMDs), goggles, eyeglasses, and mobile devices are enabled to represent virtual scenes. However, when it comes to the ultra-fast communication of 5G and 6G technology, then managing tasks on the client's side devices is a cumbersome task, as the adjustment of old and new hardware creates several problems, and Quality of Services (QoS) is among one of them \cite{bandi2022review}.  Likewise, the streaming of data to the high QoS standard with proper security protocols is another serious problem, as existing technologies are using 3G and 4G communication paradigms. Despite that, the adoption and integration of VR, AR, XR, and DT in Metaverse with the existing networks creates many problems that range from interconnectivity, communication, and QoS to security \cite{sheth2020taxonomy}. Moreover, it is also noteworthy that the existing applications using 3G and 4G communication frameworks, which fall short to satisfy the high quality of service demand in future Metaverse. To address this issue, 6G communication paradigm will incorporate several frequency bands in the electromagnetic spectrum that involves heterogeneous radios millimeter wave (mmWave) and Terahertz (THz) communication. Despite that, 6G is likely anticipated to use the non-orthogonal multiple access (NOMA) for better utilization of bandwidth capacity and spectral efficiency with the help of additional levels of power \cite{saad2019vision}. But still, we believe that the integration of present applications with emerging technologies such as AR, VR, XR, and DT in Metaverse with such an ultra-fast communication framework offers several security challenges that could be beyond the scope of present literature.

\subsection{System Resources in 5G/6G-enabled Metaverse Technologies}

\begin{figure*}[ht!]
	\centering
	\includegraphics[width=.9\linewidth, height = 10 cm]{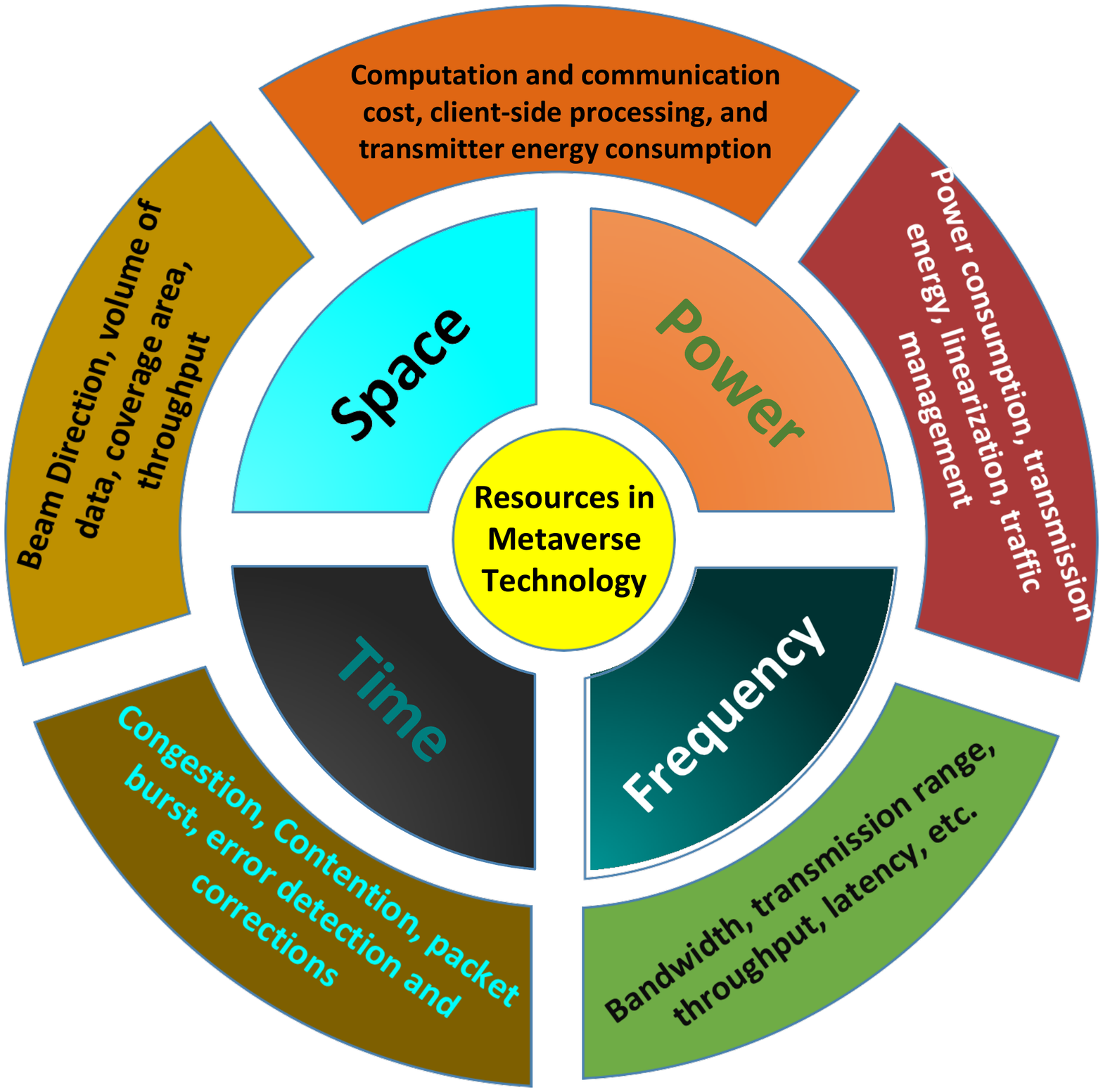}
	\caption{ Resource management in the Metaverse technology paradigm }
	\label{fig : 4}
\end{figure*}

The metaverse technology paradigm demands careful optimization of the network's available resources because this network framework consists of various components such as actuators, sensors, base stations, servers, cybers, humans, etc. To manage resources effectively among interconnected devices, different functionalities of these devices should be given considerable attention in the context of interoperability. 
Figure \ref{fig : 4} of the paper shows the resources management of the Metaverse technology in space, time, power, and frequency as a symbolic version of what is discussed in \cite{bandi2022ai, giordani2020toward}, and their resultant consequences on all communicating devices in the network followed by the services distribution, accessibility, and availability, etc.

To explore, it has been observed in the literature that fast data support is a major feature of 5G and 6G communication systems, therefore, resource allocation in the Metaverse technology with respect to time and frequency during transmission requires special consideration for reliable operation among the interconnected devices \cite{elmeadawy20196g}. Likewise, the time interlude between various periodic sensing devices' transmission beams needs efficient resource allocations to determine the burst frequency for network-friendly operation. To improve the spectrum efficiency of the 5G and 6G communication systems, the practice of beamforming had already been exercised in many technologies. With the directivity of beamforming, it has been ensured that an employed system would be able to manage a high transmission range with the embedded antenna. These tactics are likely expected to be applied to the Metaverse technology in the future.

Following this discussion, power is another very important ingredient of the system resource management for communication, sensing, transmission, and processing of data, etc. \cite{sharma2021review} With an increase in the carrier frequency, the power efficiency of a system decrease, which helps to improve the overall efficiency of the system. To explore, this area with the present literature, we suggest the readers to went through references \cite{mozumder2022overview, lim2022realizing}.

\subsection{Interoperability in 5G and 6G-enabled Metaverse}

In the Metaverse, 5G and 6G communication paradigms will aim to achieve more than 1 Tb/s data transmission. Therefore, it is pertinent to note that when it comes to the interoperability of existing technologies with groundbreaking technologies such as AR, VR, XR, and DT, etc., in Metaverse with the 5G and 6G communication frameworks. Then the present communication equipment followed by routing and security protocols would be incapable to manage the future Metaverse technologies with at least 1 Tb/s to 10 Tb/s. At present, the commercial wireless network uses a 3.4 GHz - 3.8 GHz spectrum to nourish a wide signal coverage adopted by applications. Given that, the efficient utilization of spectrum in the future Metaverse could be an unmanageable task, when it comes to the interoperability of new technologies with the existing technologies that are using traditional communication platforms.  Because emerging technologies such as AR, VR, XR, and DT, etc., will be capable to adjust with ultra-fast communication of Tb/s, while existing technologies that use 3G and 4G communication frameworks will not. As a result, many challenges will arise for the research community, and security is one them, which we discussed comprehensively in the upcoming sections, and raised several questions in the open challenges sections for the response.

\subsection{Real-Time Intelligence in 5G/6G-enabled Metaverse technologies}

In this segment, we have talked about the intelligence of the Metaverse technology. To ensure a reliable operation of this emerging technology, several challenges needed to be addressed under one umbrella that ranges from general to complex. Moreover, this emerging technology cannot be completely executed with the present IoT and WSN applications, because it requires real-time intelligence, low latency, and high bandwidth \cite{huynh2022artificial}. 
Although 5G and 6G technologies are assumed to be a high-speed communication paradigm for the Metaverse technology, but it is still going on to the emergence phase \cite{nica2022decision}.

To follow up on those challenges in the context of reliable operation, new AI-enabled techniques and algorithms must be developed. With this, it would be possible to employ 5G and 6G technology for the communication of Metaverse technology to ensure the distribution of services spontaneously and accurately \cite{kovacova2022immersive}. Keeping in view the future importance of this technology, the attackers may target different vulnerability flaws to capture the legitimate information of clients and hijack the operation of an application. Therefore, it is clearly visible that this technology will suffer new security and privacy challenges that can hamper its applicability and extend-ability in the future \cite{carter2022immersive}. Zhuo et al. \cite{zhou2022self} demonstrated that this technology offers numerous interconnectivity, interoperability, and security challenges, which need the involved stakeholders. In addition, the authors highlighted different hardware and software challenges that can be resolved with the help of AI-enabled algorithms.

\subsection{Summary of Discussion}

In this section, we explored 5G and 6G-enabled network architecture. The main reason for the inclusion of this topic in the paper was the future of Metaverse. Because, we are pretty sure that existing communication technologies should be merged with 5G and 6G technologies in the future Metaverse. Therefore, it was essential to over-view the taxonomy and communication aspects of these technologies, and set the groundwork for the main topic of this paper, which is security in 5G and 6G-enabled Metaverse. Moreover, we link the present communication technologies with the future to go through the relevant literature, and highlight the possible limitations of it. Given that, we observe these limitations to raise several open questions for the involved stakeholders. And acknowledge how these problems can be resolved in the future Metaverse.

\section{Applications of Metaverse Technology}

\begin{figure}[ht!]
	\centering
	\includegraphics[width=\linewidth]{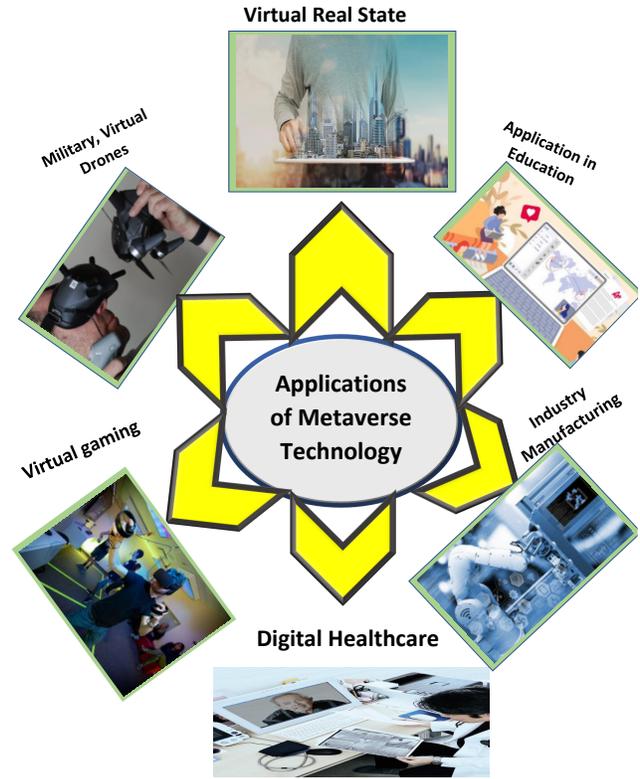}
	\caption{Diverse application domains of the Metaverse technology}
	\label{fig : 5}
\end{figure}

Metaverse technology emerged as a game-changer because it has numerous applications in many domains.  As mentioned earlier, this technology gives the idea of a virtual and shared world. Following this, leading IT players such as NVIDIA, Sony, Microsoft, Facebook, and Twitter are gambling on the Metaverse technology wonders, potential, and applications \cite{mpegv}. Bloomberg Intelligence reported and predicted that the Metaverse technology would reach almost \$800 billion by 2025 because new companies constantly looking for the factual possibility of this emerging technology \cite{kraus2022facebook}. Therefore, it is important to find the possible applications of Metaverse technology in the context of the real world. Thus, herein, we have explored the areas or application domains, where Metaverse technology had been utilized or is expected to be utilized. Different applications domains of Metaverse technology have been visually demonstrated in Figure \ref{fig : 5}. 

\subsection{Metaverse Applications in Healthcare}

When it comes to the application of Metaverse technology, the first finger goes toward the healthcare domain, because of the concerned stakeholder interest in the utilization of this technology. To exemplify, the Metaverse technology applications in the healthcare domain, the first and foremost thing we can see is augmented reality (AR) \cite{maccallum2019teacher}. AR has appeared as a potential technology to empower the base knowledge, skill, and experiments of medical students. To explore, this empowerment includes the utilization of technology-assisted surgical tools such as Microsoft Hololens that helps the surgeons in many surgical techniques \cite{mcknight2020virtual}. The adoption of such a technological procedure is the highlight of a common Metaverse application use-case that can be used in the healthcare domain to improve surgical precision and speed up the surgical procedure. 
Besides, AR headsets had also demonstrated useful contributions in MRI, 3D scan, and CT scan to facilitate the real-time data viewing of patients.  To explore this discussion in the context of Metaverse technology, it is perceptible that this technology could be very useful in the monitoring of a patient's temperature, heart rate, respiration, blood pressure, etc. \cite{tan2022metaverse}.
From the example, we have observed that the Metaverse technology points toward the use of AR in many aspects of the healthcare domain to facilitate the involved healthcare providers. Towards the imagination of virtual world transition, these examples could be assumed as a first step because this will allow healthcare specialists to check a patient from the inside precisely to figure out the problems.

\subsection{Metaverse Applications in Real Estate}

In this segment, we have talked about the applications of Metaverse technology in the context of the real state sector by taking into account VR, because VR has shown significant contributions in the recent past in different applications \cite{kozinets2022immersive}. From the operational capability of VR, it is visible that this technology can enhance the productivity of different applications, because it offers a realistic experience to the clients and application stakeholders to improve and strengthen their work and operational environment. For intense, real estate representatives may harness the potential of VR to offer virtual tours of their properties to buyers utilizing this technology \cite{dwivedi2022metaverse}. Besides, Metaverse technology also enables the integration of different multimedia features with the potential to guarantee a particular VR tour. For example, a VR real estate tour enables ambient light, music, and sound effects alongside other necessary components. Likewise, all these factors are available in Metaverse applications, which ensures real-time access to the location and information such as properties in the context of a real estate tour.  As a result of such facilitation, the real estate stakeholders can effortlessly gain the confidence of their clients by allowing them to visit different properties in real time. In addition, the clients who have access to the various features of the properties are more likely to be able to make decisions and finalize the deals.

\subsection{Metaverse Applications in Smart Cities}

Smart cities employ multifarious hardware, software, sensors, communication networks, Internet of Vehicles, (IoV), Internet of Things (IoT), and Wireless Sensor Networks (WSNs) to collect, process, and analyze data with the objective to improve the living standards of people \cite{dhelim2022edge}. Given that, this paradigm has been adopted in many countries throughout the globe to provide optimal solutions for routine problems. To exemplify, the development and deployment of technology in smart cities not only improve the efficiency of industrial manufacturing, traffic management, and agriculture sectors, but also helps to connect different applications under one network umbrella. Especially, when it comes to the evolution of Metaverse. In this technology, several other technologies such as XR, VR, AR, and MR will be integrated with the smart cities platforms to facilitate people in different aspects of life \cite{metaverse_smartcities}. Even though when we observe this transformation, it brings numerous benefits to the lives of human beings in terms of visual aspects. But when it comes to the technical side, then there are a lot of challenges that range from hardware and architecture design to application execution, which need sufficient efforts to ensure reliable operation. Among all of them, we will focus on the security concerns of this important application by keeping in view the Metaverse future.      

\subsection{Metaverse Business Applications}

At present and in the future, there are numerous business opportunities in retail for Metaverse. To explore, first and foremost, Metaverse technology provides an online and offline enterprise market platform for many business applications \cite{periyasami2022metaverse}. With such a digital framework, the business stakeholder creates an immersive shopping venture for clients. 
Given that, real-time commerce applications have been merged with the help of digital twins to overcome the limitations present online shopping technologies. With the utilization of this, users are enabled to visit stores and shopping marts in the virtual world as digital avatars to buy things just like in the real world \cite{schmitt2022metaverse}. In this platform users not only review products just like other online shopping sites but, they are enabled to check these products through digital avatars in a virtual world. With the adaptation of AR and MR in this paradigm, the business application has been further improved to facilitate customers by allowing them to understand the style and quality of products before purchase \cite{sahay2022metaverse}. Simply in Metaverse, the business trend is transformed from a "click-to-buy" practice to a new strategy known as "experience-to-buy". Despite the fact that this technology brings many things to doorstep access in the consumer market, but, at the same time, these application suffers many challenges in the context of effective technology utilization. In this paper, we will focus on the security concerns that can hamper the trust of consumer market stakeholders with their possible solutions.

\subsection{Metaverse Applications in Military}

The military sector is another important inclusion to the list of Metaverse applications. In this domain, AR and VR showcases have demonstrated remarkable potential to facilitate different applications for the sake of unique objectives \cite{jung2022study}. To exemplify this in military applications, it has been seen that tactical AR was used in the night vision system \cite{upadhyay2022metaverse}. Moreover, TAR is very useful for displaying the soldier's location alongside the allies and hostile positions.  In fact, TAR is an excellent replacement for traditional GPS and headphone devices.
This example also highlights the prospectus of Synthetic Training Environment (STE) that could be used in many military applications in the future. Specifically, this is an AR system designed for a realistic training experience for the soldiers. The STE paradigm offers a virtual environment for the soldier's training experiencing the past physical, psychological, and combat settings.

\subsection{Metaverse Applications in Industry}

In industry, Metaverse technology would have great significance and a lot of applications in the future, which can be shown from a simple search on Google, like ``Metaverse applications in the industry.'' \cite{lee2022integrated}. For instance, Metaverse applications utilizing VR and AR can be very useful in the training of employees to better understand their safety precautions alongside the prediction of risk scenarios. Similarly, the Metaverse technology can also help minimize the risk of accidents in several operational processes \cite{alpala2022smart}. Therefore, we believe that this technology can facilitate better and long-term operations with high productivity in an industrial environment. To explore this discussion with another example such as a VR headset that can be used by manufacturers to check and evaluate all the components of a product in more detail. Besides, the Metaverse applications can be useful in landscape management for the positioning of equipment and manufacturing plants \cite{guan2022extended}.

\subsection{Metaverse Applications in Gaming}

The recent progress in the gaming industry opened the door for Metaverse technology \cite{nguyen2022metachain}. With the help of this technology, people will use a virtual environment to enter the next level of real games. For better performance, gaming firms are looking for decentralized projects, because the future of this technology is decentralized \cite{jungherr2022extended}. Therefore, we would like to acknowledge the research community to stick around the virtualization of the game through Metaverse technology to help the players to understand how effectively different games could be played. To elaborate on this concept, the existing Metaverse games have a clear intention of play-to-earn, which enables the players to win the game and earn real-world money \cite{estudante2020using}. Moreover, participants can invite their social media friends to play and interact with them via Metaverse technology. 
In addition to this, Metaverse technology leverages the AR and VR paradigms to create an organic gaming environment for the players, where they feel a lifelike experience/real world \cite{shin2022actualization}. Built upon this, Metaverse technology provides an interoperable platform that allows the players to shift their game entities from one space to another space without major change.

\subsection{Metaverse Applications in Digital Twins}
In this segment, we will discuss the applications of Metaverse technology in digital twins. Digital twins, in coordination with Metaverse, is believed to be the next technological breakthrough to bridge the gap between the digital world and humans \cite{duan2021metaverse}. It has been recently noted that many stakeholders have shown interest in using Metaverse and digital twins as integrated technology because the majority of world firms accepted it \cite{zhu2022metaaid}. The most prominent domain, where the stakeholders want to use this technology includes the scanning of objects with photogrammetry, avatars creation, the manufacturing industry, the automotive industry, retailer shops, e-commerce, etc. \cite{petrigna2022metaverse}.  
Following the past outcomes of digital twins and Metaverse technology, we believed as an integrated paradigm, this technology has huge potential to enhance the productivity of the aforementioned applications. Moreover, we are pretty sure that, as an integrated technology, it could be extremely beneficial for these industries to monitor, predict, allocate, track, optimized, control, and evaluate the quality of different entities for the sake of improvement to enhance their productivity.

\subsection{Summary of Discussion}

In this section, we talked about the different applications of the Metaverse. The objective of this section was to synchronize the preceding and upcoming sections in the context of the importance of the Metaverse. Furthermore, it was pertinent to familiarize non-specialist or general readers to understand the general applicability sectors of this technology with productivity and economic contributions. Furthermore, it was also essential to let them know how this technology could useful in digital transformation or simply Metaverse. Despite that, we want to share that these not all the applications of Metaverse, because this technology has potential to extend in many domains. But herein, we discussed some of them groundwork and readers interest in this work.  


\section{Evaluation of Cybersecurity in Metaverse Technology}

\begin{figure*}[ht!]
	\centering
	\includegraphics[width=\linewidth]{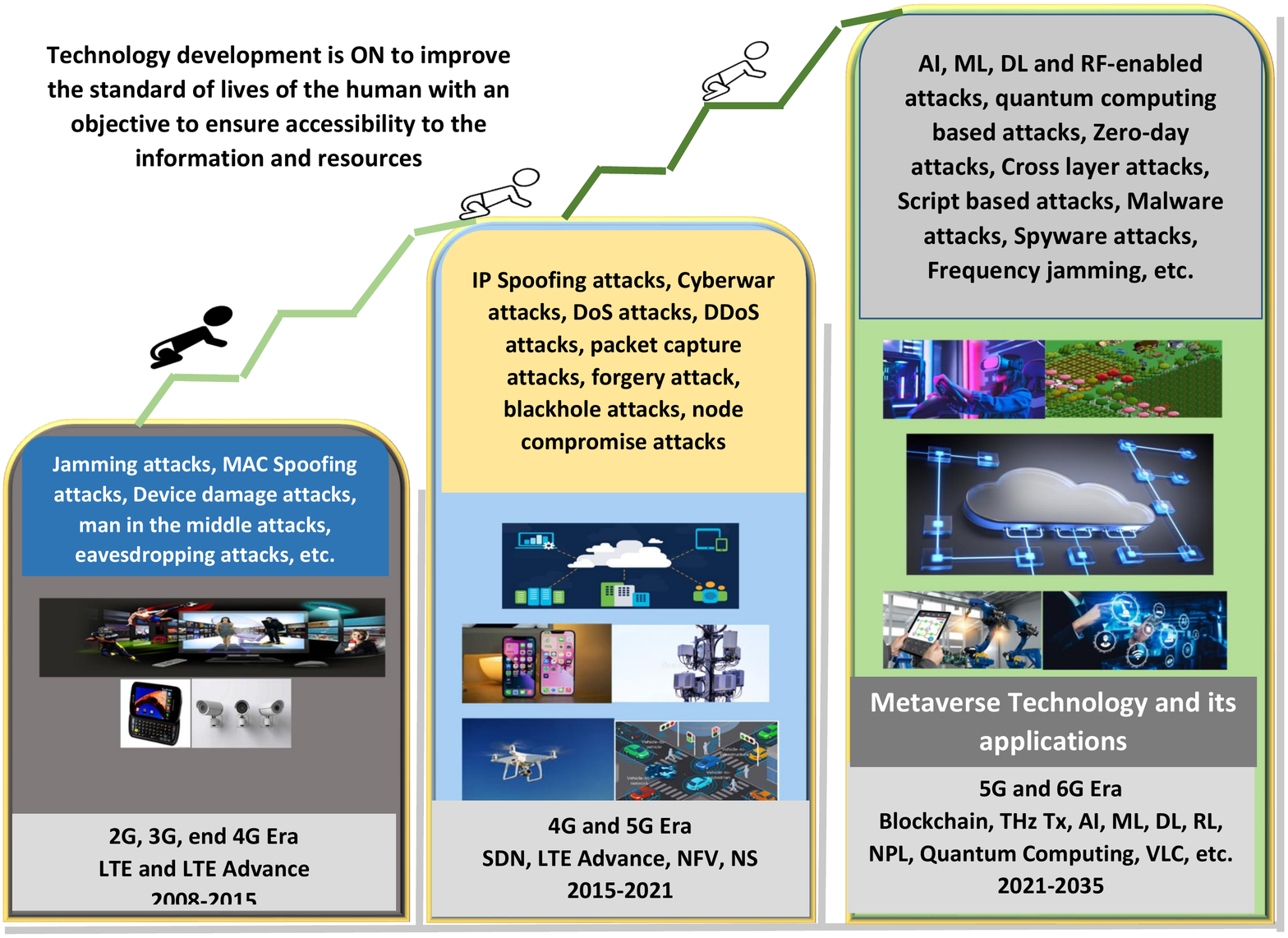}
	\caption{ Visual representation of different security threats in the context of technology progression }
	\label{fig : 6}
\end{figure*}

In this section, we intend to discuss the evaluation of cybersecurity concerns in the context of 2G, 3G, and 4G-enabled Metaverse technologies. Furthermore, we highlight different attacks such as illegal physical attacks, tampering attacks, cloning attacks, eavesdropping attacks, forgery attacks, and many more that have direct or indirect impacts on these networks \cite{lv2022blocknet}. To explore this topic more precisely, we add Figure \ref{fig : 6} to demonstrate the evaluation of security threats from 2G to 5G and 6G-enabled networks such as Metaverse. In the beginning, most of the 2G and 3G-enabled networks faced authentication and privacy threats due to wireless communication and open application execution \cite{ropero2022peer}. To exemplify, the Media Access Control (MAC) protocol was targeted at the physical layer, while the network layer protocols were targeted through IP addresses as discussed in \cite{lynch2022smart}.

Likewise, in 5G-enabled networks, most of the security threats are due to access, backhaul, wireless communication, and core networks, which are likely expected in the 6G-enabled networks as well \cite{adil2021emerging}. For 5G/6G-enabled-Metaverse technology, Cyberwarfare, Functions Virtualization (NFV), Edge/Cloud Computing, and Software-Defined Networking (SDN) technologies offer complex security challenges as highlighted earlier. For example, SDN technology had created security problems such as exposing critical Application Programming Interfaces (APIs) to unintended software \cite{seddigh2010security}. Besides, 5G and 6G-enabled Metaverse technology are envisioned with intelligent communication systems, which use advanced AI and ML-enabled routing protocols to ensure a reliable operation of concern applications. However, the integration between different technologies and algorithms in Metaverse technology knocks on the door for interoperability security challenges \cite{arfaoui2018security}.

\subsection{Security Requirements of Future 5G and 6G-enabled Metaverse Applications }

In this segment, first, we will familiarize the reader with the security requirements of  5G and 6G-enabled Metaverse technology/applications by going through the concerned literature. Subsequently, we will describe the security problems of this emerging technology by taking into account the 5G and 6G communication and network architecture. For this, we have divided these security problems into the following categories, i.e., radio core convergence, cloudification, edge intelligence, technical subnetworks management and operation. 
Before starting this discussion, we would like to acknowledge that in comparison to the existing 4G and 5G-enabled Metaverse applications, the future 6G-enabled applications would require rigid security requirements to gain the trust of involved stakeholders with reliable operations. In Table \ref{tab2}, we have summarized all the security requirements that have a direct or indirect impact on the operation, services, and performance of this emerging technology.

\begin{table*}[htbp!]
	\caption{Security requirements and challenges of 5G and 6G-enabled technologies}
	\begin{center}
		\small
		\begin{tabular}{p{1.2cm}|p{2.6cm} |p{3.3cm} |p{3.2cm}  |p{2.9cm}   |p{3.1cm}  }
			\hline
			\cline{2-4} 
			\textbf{References} & \textbf{\textit{Applications}}& \textbf{\textit{Possible Attacks}}& \textbf{\textit{Concerned Problems}} & \textbf{\textit{Communication Requirements}} & \textbf{\textit{Probability of Damage `Operation' }}\\
			
			\hline Duan et al. \cite{duan2021metaverse} 
			& Social goods application of Metaverse with University campus use case 
			& Pretexting attacks, Phishing attacks, Baiting attacks, Tailgating attacks, etc. 
			& Create distrust during communication among the professors, students, and administrative staff 
			& Delay sensitive and ultra-fast communication with packet transmission and reception rate $>$ 1 Tbps 
			& Very high: Because most of the time student responds to illegitimate text and messages.   \\
			
			\hline Zhu et al. \cite{zhu2022metaaid}
			& Industry 4.0, 5.0, and 6.0 Metaverse applications 
			& Tampering attacks, Jamming attacks, Devices Physical Damage attacks, and many more in Figure \ref{fig : 6}   
			& Disrupt the legitimate operation of Industrial employed sensors and destroy the accessibility and monitoring of these devices  
			& Delay sensitive, efficient resources management, traffic capacity about 1Gbps 
			& Extremely high, because these industries solely depend on the collected and transmitted information of interconnected devices  \\
			
			\hline Petrigna et al. \cite{petrigna2022metaverse}
			& Digital healthcare applications such as disease monitoring, patient monitoring, remote patient prescription,  and many more   
			& Jamming attacks, record tampering attacks, data preservation attacks, confidential information accessibility attacks, man-in-the-middle attacks, eavesdropping attacks, etc.
			& Can misguide doctors and nursing staff to prescribe the wrong medicine, share wrong data related to the patient's diseases, physically damage patient wearable devices, etc.
			& Authentication and data preservation with the patient mobility in the delay-sensitive communication environment, Ultra-fast up to 1 Tbps, spectrum efficiency during mobility  
			& High, because the attackers are always looking for open spaces in the network to intercept the legal communication process for the sake of different objectives \\
			
			\hline Jung et al. \cite{jung2022military}
			& Secrete military operations, Synthetic Training Environment (STE) applications for military personnel training, UAVs experimental applications, etc. 
			&  Spyware attacks, sinkhole attacks, eavesdropping attacks, IP spoofing attacks, DoS, and DDoS attacks, worm-whole attacks, MAC-spoofing attacks, Sybil attacks, forgery attacks, etc. 
			& Demonstration of immoral military drills through compromised sensor devices, observing training tactics for misuse, the misguidance of UAVs in secret operations, etc.
			& Ultra-Reliable and Low-Latency Communication (ERLLC and eURLLC) with transmission and reception rate $>$ 1, spectrum efficiency, efficient resources management, congestion-free communication, etc.    
			& Very High, because these applications are very sensitive in the context of  information sharing, time synchronization, connectivity of devices, interoperability, etc. \\
			\hline Dwivedi et al. \cite{dwivedi2022metaverse}			&  Real Estate: Properties advertisement applications, As a virtual tour enabler technology,  advertisement and tour of properties with Multimedia features, etc.   
			&  Phishing attacks,  key logger attacks, code injection attacks, social engineering attacks, eavesdropping attacks, spyware attacks, spam attacks, and many more.  
			&  Attackers through compromised devices shows different properties instead of the concerned, misguide transactions, capture the credential of clients
			& fast communication, reliable and delay-sensitive operation, and Further enhanced Mobile Broadband (FeMBB), etc.
			& Low, because of the least interest of involved stakeholders in this technology. But in the future, it could be the most targeted area.\\
			
			\hline Jungherr et al. \cite{Jungherr2022} 
			& Provides virtual gaming platform, creates social gaming network, the environment of transferable gaming,   AR and VR based gaming platform, etc.
			& Non-fungible tokens (NFTs) attacks, darkverse attacks, social engineering-based attacks, phishing attacks, spyware attacks, and attacks related to financial fraud, and many more.
			& After compromise, an attacker can steal the confidential information of the concerned client, ask for financial assets, hijack the legitimate software of your machine, etc. 
			& High traffic data such as $>$ 1Tbps, latency-sensitive, spectrum efficiency, efficient resource management, (ERLLC/eURLLC communication, etc.
			& Very high: to lose control of a machine/software, lose confidential data, lose software/application password, lose financial assets, etc.\\

			\hline
		\end{tabular}
		\label{tab2}
	\end{center}
\end{table*}

\subsection{Vision for Security of 5G and 6G-enabled Metaverse Applications}

In this sub-section, we will talk about the 5G and 6G-enabled Metaverse technologies applications, architecture, standardization, and policy implementation in the context of security concerns. Alike the general 5G and 6G-enabled applications security vision, Metaverse technology will add intelligent algorithms on the top of the network layer to softwarize and cloudify the auto-integration process of different applications with proper security protocol adoption by utilizing AI and ML techniques \cite{adil2021intelligent}. By doing this, the expert working in this domain will somehow satisfy the security concerns of the enterprise market stakeholders, but at the same time, the attacker also works restlessly to design new security hijacking techniques that could be capable of compromising the naively adopted security techniques to achieve their objectives. To explore this one step forward, it has been noted in the literature how difficult it is to identify zero-day attacks in existing 5G-enabled applications. with this, it is more challenging to stop these attacks from spreading in the network after a compromise. To tackle such challenges in newborn technology like Metaverse, the adoption of intelligent security techniques should be assumed the utmost requirement for them, because the traditional 5G-enabled applications had already demonstrated several security flaws to detect and prevent different security threats \cite{di2021metaverse, jaber2022security}.
While doing this, another important factor that needed to be considered for ensuring trust among relevant stakeholders is security and privacy to enhance the productivity of any particular application \cite{su2022survey}. Because security and privacy in a network are tightly coupled to guarantee the authentication of legitimate clients followed by the integrity of their transmitted information. In the consequent sections, we will explore the idea of how security and privacy are correlated to each other in the context of 5G and 6G-enabled Metaverse technologies/applications.

\subsection{Key-based Security Requirements of 5G and 6G-enabled Metaverse Applications}

In this segment, we talk about the importance of Key-Value indicators (KVI) and Key Performance Indicators (KPIs) to set the stage for its utilization in the 5G and 6G-enabled Metaverse applications. In the future, it is likely expected that 5G and 6G-enabled Metaverse applications will integrate technologies such as smart sensing, artificial intelligence, edge computing cloud computing, embedded devices, machine learning algorithms, and many more as discussed in \cite{hussain2019integration, adil2020anonymous}.
Following the future expectations of this technology, the role of KPIs and KVIs can not be neglected in the context of authentication of legal devices, because there will be some static and mobile networks that would be connected under one heterogeneous network paradigm. For this, the authentication accuracy, computational time such as a handshake, and AI model convergence time could be of great importance because most of the existing KPIs schemes do not satisfy this requirement in 5G and 6G-enabled applications \cite{adil2021three}.
According to the UN sustainable development goals (SDG) for sustainability, accessibility, security, and trustworthiness of Metaverse technology, the KPIs and KVIs could be used as promising security technology to determine the values of the new 5G and 6G-enabled Metaverse technologies \cite{rehm2015metaverse}. Therefore, we are pretty sure that the new characteristics of KPIs and KVIs will have a substantial role in improving the security of Metaverse technology.

\subsection{Security Threats to 5G and 6G-enabled Metaverse Applications}

\begin{figure}[ht!]
	\centering
	\includegraphics[width=.95\linewidth]{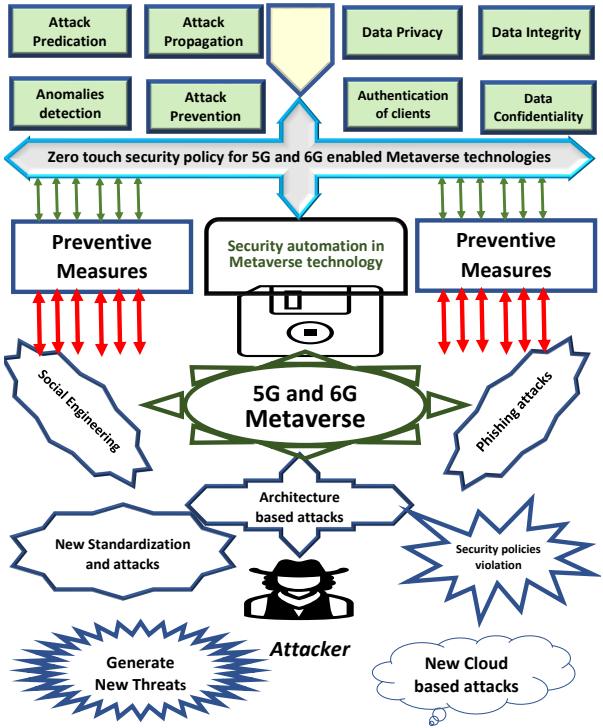}
	\caption{Possible security threats and their counteractions to 5G and 6G-enabled Metaverse technology applications }
	\label{fig : 7}
\end{figure}

Undoubtedly, in the future, the enormous growth of 5G and 6G-enabled Metaverse technology will increase the risk of vulnerabilities in different applications in the context of security and privacy \cite{uusitalo2021hexa}. To familiarize the readers and researchers with the possible security threats, we have added Figure \ref{fig : 7} and shown how attackers launch new attacks and what we can do as an expert to counter these attacks. Specifically, we have highlighted the possible attack areas at the bottom of the figure, whereas at the top, we have shown what we need to do as a counteraction. 
Moreover, we have considered the four core areas such as anticipated technologies, network architecture, integration of environment, and application-specific requirement. For the non-expert, we have added textual explanations to enhance the worth and readability of this work. To more realistically consider the security threats of this emerging technology, we have taken into account the Nokia Bell Labs use case as an ambitious approach for the connectivity of this technology to support our arguments.

Following the core-concept security discussions, herein, we would like to start with data architecture security.  This part is further classified into sub-categories such as functions, platform, specialization, and orchestration. Based on this classification, we can see that the heterogeneous cloud framework for Metaverse technology could be developed in a fashion that will be open, expandable, interoperable, and agnostic with 5G and 6G-enabled networking entities to speed up the communication process with secure data flow in the network \cite{yrjola2022value}. For exemplification of this scenario, we can consider the intelligent radio and RAN convergence for \textbf{``functional''} architecture \cite{bi2021improved}. Likewise, the \textbf{``specialized segment''} deals with resource management, packet slicing, and sub-networking in these technologies, while the \textbf{``orchestration''} part handles the problems like loop automation and intelligent network management.
Although with the innovations in technologies, the security frameworks have been improved by utilizing new algorithms to mitigate different attacks. But, the attackers are working with the same potential to hijack them and get unauthorized access to the network. Therefore, we would like to acknowledge the concerned stakeholders to be vigilant about newly adopted attack mitigation techniques.

\subsubsection{Intelligence and Convergence Requirements for Security in the Radio and RNA} \hfill

The recent advances in antenna circuits and meta-material in coordination with machine learning algorithms showed spectacular results in overcoming problems encountered by wireless communication in the past \cite{rahimi2018security}. 
To further improve this infrastructure for naive technologies such as \textbf{Metaverse} in the future, it is required to use intelligent cognitive radio systems (CRS) in coordination with machine learning algorithms \cite{mahmood2021industrial}. In the recent past, CRS technology had shown significant contributions to emerging technologies like the Internet of Things, the Internet of Vehicles, Unnamed Aerial Vehicles, etc. \cite{noohani2020review} From the past results, it is clearly visible that this technology could be a game changer if utilized with edge-cutting AI/ML algorithms to manage the security concerns related to channel modulation, resource allocations, spectrum access, beamforming, etc., of the Metaverse technology applications. Moreover, the induction of this technology will also reduce the hardware cost and implementation time with better communication metrics, which is the basic requirement of Metaverse technology applications. For instance, AI/ML model will manipulate the spectrum access system by inserting fake signals to diverge the attackers' request from network anticipation.

\subsubsection{Edge Intelligence Security Requirements in Metaverse Technologies } \hfill

In Metaverse technologies, the intersection between edge computing and AI/ML algorithms is spontaneous, because they work together for one objective \cite{yoon2022rna}. In the existing literature, it has been outlined that many 5G and 6G-enabled Metaverse applications highly rely on the computing of edge-employed devices for better operation and results \cite{guo2022artificial}. From this, we can simply define edge intelligence as the collective processes of edge devices and AI/ML algorithms to collect, store, and transmit information in the network \cite{xu2022metaverse}. 
Specifically, in this integrated environment, AI/ML model analyzes data for anomaly detection before forwarding it to the servers for further processing in the network, which helps to improve the latency of an employed network. Despite this advantage, edge devices are still proven to be vulnerable to several security threats, because of high data dependencies. Following these constraints, edge devices need reliable AI/ML-enabled authentication, data preservation, and access control models that could be capable of ensuring the integrity and confidentiality of information during transmission.

\subsubsection{Security Requirements of 5G and 6G-enabled specialized Metaverse Applications } \hfill

In the literature on Metaverse technology applications, we have noted that 5G and 6G-enabled industrial automation will work as a sub-network to this heterogeneous network platform \cite{chang20226g}. Likewise, this specialized network is expected to work in a stand-alone platform, where several applications would be connected together to perform different tasks. In contrast to its wonderful advantages, the wireless connectivity of these subnetworks exposes them to several security threats, which creates distrust among the associated stakeholders \cite{cao2022decentralized}. To avoid these problems, it is pertinent to use strong and lightweight authentication protocols that could be capable of ensuring the security of clients and vendors.
In the literature, different authentication schemes have been used to resolve the security concerns of edge and cloud devices in the Metaverse technologies; but, somehow, they are computationally complex and communication expensive \cite{kechadi2022edge}. Therefore, lightweight authentication and data preservation schemes should be designed in the context of the dynamic and hierarchical model to set the security boundaries for subnetworks cost-effectively. Besides, the Trusted Execution Environments (TEE) should be used as an alternative promising security technology to address the data confidentiality, availability, and integrity \cite{wei2022gemiverse}.

\subsection{Security Requirements During Interoperability and Network Management } \hfill

Extreme range and high power transmission of 5G and 6G-enabled Metaverse technologies demand low latency and a cost-effective communication environment for interconnected devices such as machine-to-machine, machine-to-toys, and machine-to-human, etc., with a proper security framework \cite{peukert2022metaverse}. In the literature evaluation, we have noted that the existing authentication and data preservation schemes would not be capable of managing the security of future 5G and 6G-enabled Metaverse technologies, due to ultra-fast communications among interconnected entities in the network \cite{salameh2022from, caison2022relationship, klotins2021siot, jaiswal2017security}. With the help of AI/ML algorithms, these technologies are expected to offer reliable operations with a secure network management framework \cite{haddadpajouh2021survey, hameed2019understanding}. Therefore, it is very important to discuss these requirements that can directly or indirectly hinder the operation and performance of this emerging technology. Herein, we have added Table \ref{tab3} to highlight different attack domains in the context of Metaverse technology. In addition, we will also underscore the possible solutions for the redressal of accentuated attacks.

\begin{table*}[htbp]
	\caption{Security attacks and their possible solutions associated with the network interoperability and management}
	\begin{center}
		\small
		\begin{tabular}{p{2cm}|p{2.5cm} |p{4.5cm} |p{3cm} |p{2cm} |p{2.2cm}  }
			\hline
			\cline{2-4} 
			\textbf{Possible Attack Area} & \textbf{\textit{ Attacks name}}& \textbf{\textit{Description of Threats}}& \textbf{\textit{Possible Solutions}} & \textbf{\textit{References}}  & Percentage \\
			\hline Security Threats related to open API
			& Identity-based attacks, Parameter based attacks, Hardware based attacks
			& During interoperability, cross-domain validation flaws offer parameters-based attacks. Malicious data injection during the network manipulation phase, False parameters configuration of hardware, etc. 
			& Client authentication and validation, Input data verification, network access control policy, and buffer management
			& Garg et al. \cite{garg2019securing}, Harsha et al. \cite{harsha2018analysis},  Rafiq et al. \cite{rafiq2022mitigating}
			& High possibility of attacks ranges from 65 to 80 \% \\
			\hline Intent-based attacks on Interfaces 			
			& Traffic Behavior based attacks, Information exposure attacks, undesirable configuration
			& Misguides the processing commends from intent to actions, Tamper the security priorities such as changing its position from High to Low or Low to High  
			& Intent mapping verification and validation, clients authentication with respect to input data, Traffic analysis via AI-enabled algorithms,
			& Kim et al. \cite{kim2020ibcs}, Hyder et al. \cite{hyder2020inmtd}, Klymash et al. \cite{klymash2021future}, Safavat et al. \cite{safavat2020elliptic} 
			& Probability of attacks medium ranges from 50 to 70\% \\
			
			\hline Network in loop attacks on automation processes
			
			& Forgery message attacks, fake rout advertisement attacks, fake event advertisement attacks
			& Generate high traffic load in the network via fake requests, dodge legitimate devices via fake authentication requests, and disrupt the communication process through fake event advertisements. 
			& New data encryption and decryption schemes, Use of VPN and other secure routing protocols such as IPSec, SSL, and TLS, etc.
			& Verma et al. \cite{verma2020security}, A. Almusaylim et al. \cite{almusaylim2020detection}, Seyedi et al. \cite{seyedi2020niashpt}
			& Probability of attacks is very high ranging from 70 to 95\%\\
			
			\hline Semantic Attacks on Metaverse Technology
			
			& defamation attacks, SQL-injection attacks, Botnet attacks, and poisoning attacks 
			& In these types of attacks, an attacker focus on legitimate data alteration in such a way to misguide the end-side administrators and the client's conclusion.
			& Cryptography-based solutions, software-based solutions, routing protocol-based solutions, SSL, IP Sec, etc. 
			& Gowtham et al. \cite{gowtham2021semantic}, Rethinaval et al. \cite{rethinavalli2022botnet}, Yan et al. \cite{yang2021smartdetour}
			& Probability of attacks is high, and ranges from 70 to 85\% \\
			
			\hline Syntactic attacks on Metaverse Applications
			& Trojans attacks, spam attacks, worms attacks, and viruses attacks
			& In syntactic attacks, an attacker uses different types of viruses, spam messages, scripts, and software pieces to penetrate legitimate devices and hijack their security.
			& Message identity-based solutions, anti-virus-based solutions, auto-software update-based solutions
			& Lüdtke et al. \cite{ludtke2021attack},  Luo et al. \cite{luo2021vulnerability},  Pashamokhta et al. \cite{pashamokhtari2022adiotack}
			&  Probability of attacks on open area employed networks is very high and ranges from 75 to 95\ \\

			\hline
		\end{tabular}
		\label{tab3}
	\end{center}
\end{table*}

\subsection{Consumer Side security Requirements of 5G and 6G-enabled Metaverse Technologies}

In the early generation of Metaverse applications, secure data transmission depended on the physical insertion of symmetric keys in the edge-/network-employed devices subscriber identity module (SIM), which was an arduous task \cite{almasarani20215g}. Later on, cryptographic tools had been used to ensure the authentication and verification of network-connected devices, but they used computationally complex techniques for this task, which limited their deployment in many resource-constraint networks \cite{di2021increase}.  
According to the literature and general standards, 5G-enabled communication entities are still using SIM cards for their security verification, which is a practical example of inserted keys module. Therefore, it is likely expected that this may limit the predicted growth of 5G and 6G-enabled Metaverse technology applications in the context of security challenges \cite{silva2021esim}. To handle this problem effectively in the future 5G and 6G-enabled Metaverse technologies, we suggest the concerned stakeholders check the feasibility of e-SIM cards. 
For the solution of this problem, we also suggest the utilization of iSIM that could be adopted in the future gadget's hardware as a system-on-chip module (SCM). However, this will offer some challenges for the telecom sectors in the context of conceivable loss of control, which needs considerable research at the moment. In contrast, SIM cards use symmetric key encryption techniques, which allow them to scale up to millions and billions of devices. 

Keeping in view these facts, 5G and 6G-enabled Metaverse technologies need a remarkable shift from symmetric cryptography to asymmetric cryptography or even to post-quantum cryptography. Following the future plan of 5G and 6G-enabled technologies, it is clearly visible in the literature that they will use public-key infrastructure (PKI) for authentication and secure communication in the network. 

\subsection{Security Requirements of AR in Metaverse}

AR technologies are comprised of sensors, cameras, displays, goggles, eyeglasses, earpieces, microphones, etc. These devices need efficient utilization for better operation and productivity \cite{adil2021ai}. Despite these factors, it is challenging to ensure the security of these devices with different customers, because it is likely expected that in Metaverse, these gadgets will be used by different clients \cite{kurt2022security}. Following this, it is also hard to maintain the security parameters of every individual using these devices, because it has limited memory. Despite that, it is needed to understand, when the authentication parameters of new users are processed through these devices, then how the computation capability of these devices should be maintained \cite{alismail2022systematic}. These are very important factors that required the attention of the involved stakeholders.  Furthermore, we will discuss the open security questions with this segment in the concerned section.

\subsection{Security Requirements of VR in Metaverse}

Since 2016, VR technology has shown a considerable contribution to the general public in many applications with good experience. To exemplify, a head-mounted display (HMD), sensors, cameras, earpieces, microphones, etc, are used as individual technology or coordinated technology to perform different tasks in Metaverse \cite{zhernova2022overview}. Despite the contributions, these technologies suffer several problems in terms of communication metrics and security. Therefore, it is important to know when VR would be integrated with the existing or old network applications in Metaverse. Then how the security parameters of aforestated gadgets and current network entities should be made compatible? Because these gadgets have limited resources, whereas the existing applications use hard authentication parameters such as default key matching, two factors or three factors authentications, etc, which is beyond the scope of the aforementioned devices. Therefore, it is essential to design lightweight authentication frameworks for future Metaverse. Furthermore, we will explore this topic in the open security challenges, because the present literature suggests traditional authentication and data preservation schemes, while VR technology demands smart and lightweight authentication and data preservation schemes.

\subsection{Security Requirements of XR in Metaverse}

In traditional applications, security is a well-established field. But in Metaverse and particularly XR, this field still needs a lot of attention and groundwork, because of the HCI followed by the real world and virtual world transition \cite{syed2022car}. Even though some work is in progress, but there are no sufficient results for them, because of the complex paradigm of Metaverse. Moreover, conventional security techniques are not useful for this technology, as there are too many factors, which need to be considered while designing an adversary-proven security framework for them. These factors include but are not limited to client security, gadget security, application security, data security, and site security, etc.   Therefore, it is necessary to take care all of the aforesaid factors, while devising a security framework for this emerging technology.  

\subsection{Security Requirements of Digital Twins in Metaverse}

In Metaverse, a digital twin technology is used to generate virtual copies of objects, systems, shops, and even the world in real-time with accurate representation \cite{masaracchia2022digital}. Although this technology has shown remarkable assistance in digital transformations. But at the same time, it has demonstrated some real challenges for associated people. One among them is security, because this technology needs special security techniques instead of conventional security models \cite{alcaraz2022digital}. Therefore, we want to mention the research community needs to think about physical objects followed by virtual objects, while designing the new security countermeasure schemes. Despite that, they also need to take care of the application requirements accompanied by network capability, as the future Metaverse will use 5G and 6G communication paradigms.

\subsection{Summary of Discussion}

In this segment, we talked about the security requirement of 5G and 6G-enabled Metaverse technology applications. With this, we demonstrated that how the security requirements of Metaverse are different from the traditional network applications. Furthermore, we discussed different components of the Metaverse with their operational capabilities. The basic notation behind this was to familiarize the readers and researchers with the requirements of this technology by keeping in view the resource-limited objects followed by the integration of emerging technologies such AR, VR, MR, XR, and DT, etc. From this, it is clearly observable that the resources of employed Metaverse objects need to be utilized efficiently for better operation and results. Therefore, we strongly believe that the aforementioned metrics should be given considerable importance while designing new authentication schemes or modifying the existing authentication schemes for future Metaverse technology applications. Before doing this, we would like to go through the present literature in the upcoming sections to familiarize the readers with the current work. In addition, we will highlight the limitations of different schemes, which could be assumed to be a step forward for the open research challenges and future research directions.

\section{Security Challenges with Existing Literature}

\begin{figure*}[ht!]
	\centering
	\includegraphics[width=.95\linewidth, height = 12 cm]{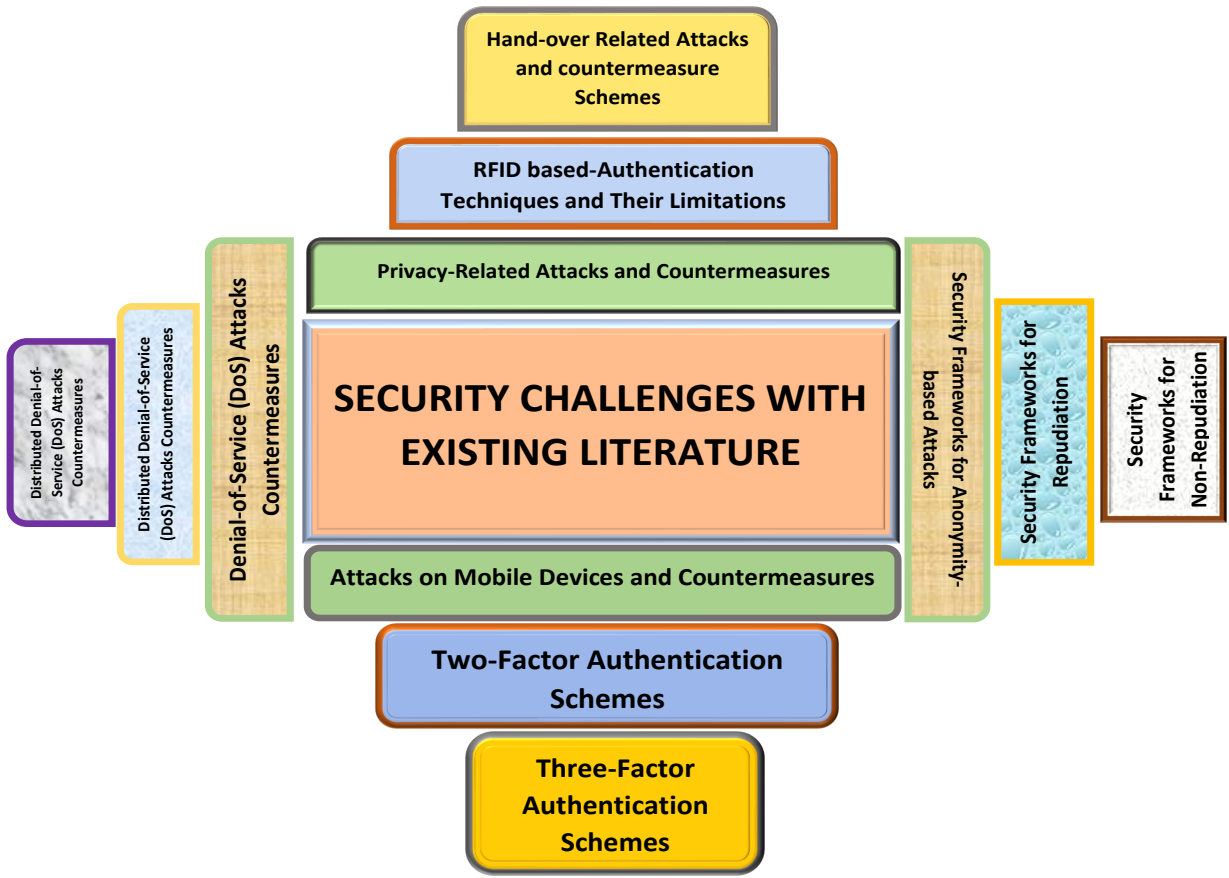}
	\caption{Present Security challenges and their counteraction schemes visual representation}
	\label{fig : 9}
\end{figure*}

In this section, we will discuss the existing literature associated with the security of Metaverse technology by taking into account the aforementioned security requirements and different attacks. With this, we will highlight different attack scenarios to underscore the limitations of the present literature. For visualization, we added figure \ref{fig : 9} in the paper.

Utilizing these problems, in the open security challenges sections, we will examine the open research questions associated with this technology that still need the engaged stakeholder's attention.

\subsubsection{ Privacy-Related Attacks and Countermeasures  }

In this section, we have considered different attacks, which can hamper directly or indirectly the integrity and privacy of transmitted data in these networks. Moreover, we have followed the preceding sections to give a brief overview to readers regarding different attacks and their counteractions. Following this discussion, we know that from the emergence of technologies such as Metaverse, it is clearly visible that the world is moving fast toward digital transformation. But, security susceptibility like this will create a problem for involved stakeholders. Despite the security concerns, this technology would be very helpful to facilitate humanity in different aspects of life, but these problems put them at risk as discussed in \cite{huang2020secure}. Such risks need to be tackled through reliable data privacy and preservation techniques to maintain the trust of all associated entities. For this redressal, Almusaylim et al. \cite{almusaylim2018proposing} proposed a hybrid machine learning-enabled protocol by considering the limitations of the Lewis et al. protocol.  

A two-step data preservation model was proposed by Zhang et al. \cite{zhang2020edge} to resolve the security-related problems of wireless networks. In the first step, the authors used a fuzzy logic-based mathematical model to confirm the selection process of participating devices, while in the second phase, they used a key-matching approach to guarantee the authentication of legitimate devices. Kang et al. \cite{kang2022blockchain} proposed  a cross-chain-enabled federated learning framework for Metaverse IoT applications to improve the integrity of data during transmission. In this model, the authors used two different, but interconnected frameworks such as physical and virtual blockchains to train the suggested model and check its result performance. Gong et al. \cite{gong2021data} continued this discussion and proposed a ring signature based-proxy re-encryption scheme for blockchain-enabled networks to ensure their security requirements by keeping in view the Metaverse technology future. Through this scheme, the transmitted data was stored in the system as a cyphertext until its verification and validation process completion, because the author's objective was to minimize the computational cost and time during this phase. Wang et al. \cite{wang2021privacy} used a cloud-based security model named ``privacy-enhanced retrieval technology (PERT)'' to ensure the integrity of data during the communication process. The author used an information-hiding technique to preserve data privacy during the edge server and cloud communication process. Xiong et al. \cite{xiong2019ai} proposed an artificial intelligence (AI)-enabled three-party game (ATG) framework by combining a machine learning algorithm with game theory to ensure the privacy and integrity of data in distinct wireless network applications. To elaborate, the authors used a random forest classifier with a K-anonymity algorithm to design the classification model. Thereafter, they used a three-party game model to analyze the raw data on the client side for different anonymity detection. Li et al. \cite{li2018anonymous} suggested a novel anonymity-based privacy-preserving data collection (PPDC) for healthcare IoT applications to address the security problems of these networks. The authors used the (a,k)-anonymity framework to analyze the client-server traffic for different attack scenarios.  To further explore this area, interested researchers could refer to the recommended survey articles \cite{chanal2020security, sarwar2018brief, dalipi2016security}.

\subsubsection{RFID based-Authentication Techniques and Their Limitations}

In this section, we will discuss different RFID-based authentication and data privacy schemes that had been used in the recent past to secure Metaverse technology applications. In this regard, Jangirala et al. \cite{jangirala2019designing} proposed a lightweight radio frequency identification (RFID)-based authentication protocol for blockchain-enabled 5G mobile computing networks to resolve the authentication, verification, and validation problems of these networks. To minimize the computational complexity on the client side, the author used a one-way cryptographic hash function with rotating bitwise operations in the proposed model. Likewise, Ahmed et al. \cite{ahmed2021cloud} proposed a cloud enable-RFID authentication scheme for smart cities IoT networks by taking into account the smart homes case study scenarios. 
Zhang et al. \cite{zhang2021privacy} suggested a privacy-preserving contact tracing-based technique to mitigate different security threats of 5G-enabled healthcare IoT applications. This model was checked against several attacks for reliability and comparative metrics. Although this scheme was resilient against considered attacks, it was computationally expensive in terms of authentication and verification request processes. Gope et al. \cite{gope2018lightweight} also used RFID for the security of IoT applications by considering the future smart city infrastructure. In this work, the authors considered the forward secrecy, untraceability, and anonymity of RFID tags to ensure the legitimacy of interconnected devices in the network. In \cite{gope2021provably}, the authors proposed an anonymous authentication of Physically Unclonable Functions (PUF) for RFID-enabled UAV-assisted IoT applications.

Kumar et al. \cite{kumar2020rseap} proposed an elliptic curve cryptography-based authentication model for wireless network applications utilizing the RFID framework. For result analysis, this scheme was checked against reply attacks and man-in-the-middle attacks. However, the authors did not provide any proof regarding the resilience of forgery attacks, wormhole and blackhole attacks, etc., which hinders the implementation of this protocol in the real environment. Soni et al. \cite{soni2022rfid} proposed an intelligent RFID-enabled security framework for healthcare applications of WSNs and IoTs to counter DoS attacks. Later on, some researchers identified that this model was effective against DoS attacks, but unable to counter several relevant attacks, which limited its use in the real environment. In \cite{yusoff2021role}, Yusoff et al. presented a comprehensive survey regarding different RFID security techniques that had been used in the past to counter the existing threats of WSN and IoT applications. Moreover, they also acknowledged that these techniques would be useful in the future 5G and 6G-enabled Metaverse technology. 

\subsubsection{Authentication in Mobile Applications}

Metaverse technology has been emerging as a heterogeneous network of networks, where many applications of mobile networks, flying networks, and static networks would be connected in one topology order. Therefore, it is important to go through the present literature that is associated with different entities' authentication participating in the network. By doing this, we will highlight the pros and cons of related studies that could be used to examine the open research challenges connected with this topic.    

Vishwakarma et al. \cite{vishwakarma2021scab} proposed a security framework for blockchain-enabled IoT applications known as SCAB-IoT (security communication and authentication for blockchain). In this model, the authors used advanced encryption standards (AES) in coordination with the elliptic curve digital signature algorithm (ECDA) to ensure the authentication of network devices followed by the privacy of transmitted data. In \cite{deebak2022tab}, the authors proposed a trust-aware blockchain-enabled seamless authentication protocol to resolve the authentication and data privacy issues in IoT applications by keeping in view their future plans in the context of integration with other applications. The authors worked on the traffic pattern of the interconnected devices in convergence with authentication processes to improve the communication metrics such as packet lost ratio and latency. Prakasam et al. \cite{prakasam2022low} proposed a low latency, area, and optimal power Hybrid Lightweight Cryptography Authentication Scheme utilizing the 8-bit manipulation principle to ensure the verification and validation of legitimate devices in the networks. In \cite{satamraju2021decentralized}, the authors suggested a decentralized authentication scheme for blockchain-enabled IoT applications utilizing the Physical Unclonable Functions (PUFs) of participating devices. In this model, the author focused on the computational complexity during the validation phase to improve the lifetime of resource-limited devices. Masud et al. \cite{masud2021lightweight} presented a lightweight and anonymity-preserving user authentication protocol for IoT applications to counter different security threats of these networks. 

\subsubsection{Two-Factor Authentication Schemes}

In the recent past, different authentication techniques have been used to ensure the security of IoT and WSN applications by preserving the future integration of these applications with heterogeneous networks such as the Metaverse technology paradigm. Two-factor authentication framework is one of those strategies that have demonstrated satisfactory results. To exemplify, Kariapper et al. \cite{kariapper2021attendance} proposed a two-factor approach in collaboration with RFID to ensure the security of industrial IoT applications. In this model, the authors used a microcontroller, Global System for Mobile Communications model (GSM), and an RFID tag to verify the legitimacy of interconnected machine-embedded sensor devices. In the second phase, they used a recording camera that worked in coordination with a multi-task Cascaded Convolutional Network to validate the legitimate devices. Likewise, when both phases satisfy the security parameters in terms of matching, then they are allowed to communicate in the network. 
In \cite{hossain2021icas}, the author proposed a two-factor authentication scheme for future integrated wireless networks. This model was effective for client-server authentication because the authors only emphasized on one-hop communications. In multi-hop communications, this model could not be applied because in this scheme the server acts as a trusted authority, which still leaves device-to-device as an open challenge. Kaur et al. \cite{kaur2021cryptanalysis} suggested an enhanced two-factor authentication for future 5G-enabled networks by considering password-guessing attacks, reply attacks, and gateway bypass attacks. 
Sadri et al. \cite{sadri2021anonymous} proposed an anonymous two-factor authentication scheme for wireless sensor networks by assuming different attack scenarios. Furthermore, the authors assumed communication and computation metrics for comparative examination to proclaim the significance of this scheme in the presence of rival schemes. 

\subsubsection{Thee-Factor Authentication Schemes}

To address the authentication problems in integrated wireless networks, Abdi Nasib Far et al. \cite{abdi2021laptas} proposed a three-factor authentication scheme for these networks utilizing MasterCards.  Amintoos et al. \cite{amintoosi2021tama} suggested a three-factor authentication model of the elliptic curve cryptography algorithm (ECC) to ensure the security of interconnected entities in wireless networks. In this model, the author tried to minimize the authentication complexity on the client side. Jiang et al. \cite{jiang2021three} presented a three-factor authentication scheme for future wireless networks by taking into account the physical Unclonable Function (PUF) with the AKE protocol. Specifically, the authors used PUF, fingerprint, and password in collaboration for the authentication and validation of legitimate devices in the network. Chen et al. \cite{chen2022novel} presented a novel biometric three-factor-based authentication scheme for 5G-enabled future IoT applications utilizing a multi-server infrastructure. 
In \cite{cho2022secure}, the authors suggested a secure three-factor mutual authentication protocol to ensure the security of e-governance systems using a multi-server network framework. For formal security analysis, the author used AVISPA (Automated Validation of Internet Security Protocols and Applications) simulation tool with Burrows-Abadi-Needham (BAN) logic, and Real-or-Random (ROR) model to verify the effectiveness of this scheme against different security threats. Following this discussion, Meshram et al. \cite{meshram2021efficient} proposed a three-factor authentication scheme of biometrics, smart cards, and passwords for IoT applications by considering their wireless communication environments. This model was checked in the AVISPA simulation tool for formal security analysis against different types of attacks. To explore this topic furthermore, we suggested the readers refer to a review article by Chuang et al. \cite{chuang2021independent}, in which the authors evaluated 49 three-factor authentication schemes that had been used between 2013-2020 to ensure the security of wireless networks and their applications. In addition, the authors also discussed the pros and cons of these schemes to provide a path for new research in this domain.

\subsubsection{Denial-of-Service (DoS) Attacks Countermeasures}

In the recent past, deniable authentication has been used as a promising authentication technique to counter different attacks associated with the network and session layer of the network \cite{huang2019authentication}. In contrast to traditional authentication schemes, this technique does not use a third party for the verification and validation of legitimate devices in the network. Therefore, they are very effective and helpful in resolving the authentication problems on the client side of the future 5G and 6G-enabled wireless network applications. To exemplify, Ibtissam et al. \cite{ibtissam2022assessment} proposed a testbed to check and verify denial-of-service (DoS) attacks on the hardware side of sensor devices. For the implementation and justification of this model, the authors presented a case study of a sensor transmitter. In \cite{olakanmi2021throttle}, the authors proposed a time-tracking-based digital signature scheme for future wireless networks to avoid DoS attacks. During simulation, the proposed scheme was checked against several attack scenarios and showed satisfactory results. Chunka et al. \cite{chunka2022secure} proposed a secure Key agreement protocol known as Defiant to counter DoS attacks in 5G and 6G-enabled networks. This model was checked against DoS and DDoS in the simulation environment to verify its effectiveness in terms of attack detection and prevention. In \cite{lebepe2021evaluation}, the authors proposed an integrated security paradigm of Software-Defined Networks (SDN) and Cognitive Radio Networks (CRN) to counter DoS attacks in future 5G and 6G-enabled networks.

\subsubsection{Security Frameworks for Anonymity-based Attacks}

In this segment, we will review literature related to anonymity-based attacks to familiarize the reader with this devastating attack type. Following this, we will also include relevant literature to discuss the existing countermeasure schemes and their limitations. Based on that we will underscore the open security challenges associated with anonymity attacks. Saleem et al. \cite{saleem2022anonymity} explored this topic by presenting a comprehensive survey related to this topic. In this paper, the authors discussed different counteraction schemes that had been used in the recent past to mitigate anonymity attacks in wireless networks.  Dwivedi et al. \cite{dwivedi2021multi} introduced an intelligent multi-parallel adaptive evolutionary technique to counter anonymity attacks in IoT applications. In this model, the authors used the grasshopper optimization technique to improve the attack detection ratio in operational networks. 
Rasheed et al. \cite{rasheed2021exploiting} used the zero-knowledge proof (ZKP) technique to detect and prevent anonymity attacks in future 5G-enabled IoT applications. In the model, the authors ensured the security of IoT devices through the newly designed multimode ZKP protocol by taking into different attack scenarios. In \cite{zola2022attacking} the authors proposed a GAN-enabled technique for anonymity detection in the blockchain networks of bitcoin. In this model, the authors classified the network traffic based on their behavior to detect illegal requests/traffic in the network. 
In \cite{martins2022host} the authors presented a detailed review of anomaly detection through Intrusion Detection Systems by discussing the real dataset performance in different adopted techniques. Moreover, the authors also highlighted some publicly available datasets that can be used for the evaluation of anomaly detection in real networks. 
Following this discussion, Chen et al. \cite{chen2022improved} proposed an advanced clustering-based anomaly detection framework for next-generation networks. In this model, the authors used a novel density peak clustering algorithm to screen out the grid's internal process followed by the network traffic for anomaly detection. 
In \cite{krim2022adaptive}, the authors introduced an adaptive threshold-based security framework for ATMs to resolve the anomaly detection issue in these networks. As for this model, the authors claimed that it is flexible, and has the capability of scalability, therefore, it could be implemented in other relevant domains.

\subsubsection{Security Frameworks for Repudiation and Non-Repudiation-based Attacks}

In this part, we talked about the existing literature associated with the Repudiation and non-repudiation attacks of the Metaverse technology applications and their counteraction schemes. 
In \cite{divya2022nonrepudiation}, the authors presented a detailed survey regarding the non-repudiation attacks on IoT applications. In this paper, the author only followed the non-repudiation counteraction schemes to highlight its limitation. Although this was a good paper, but the author did not demonstrate the open security challenges accompanied by future research directions, which makes this work worthless. Because, it does not provide a complete follow-through to readers, students, and experts working in this domain. Chen et al. \cite{chen2022trustbuilder}, proposed an integrated security framework for blockchain-enabled IoT applications to counter repudiation and non-repudiation attacks. For encryption of data, the author used cloud services, while for edge device security, they used a one-way hash digest authentication framework. This model was very useful in terms of security management, but its high computation cost does not allow its applicability in the real world. 
In \cite{wang2022ttracer}, the author introduced an advanced encoding-based-watermarking scheme known as "T-Tracer" to handle the repudiation and non-repudiation attacks in real networks. In this model, the author bound each user for a digital fingerprint before transmitting any data in the network. 

Sharaf et al. \cite{sharaf2022nonrepudiation} extended this discussion and proposed a non-repudiation private membership test (NR-PMT) model to mitigate non-repudiation attacks in electronic healthcare records. In this model, the authors used cryptographic techniques in the coordination of key exchange to ensure the privacy of transmitted data in the network. For further interest in this topic and in-depth analysis, we suggest the following survey articles \cite{ayele2021non }.

\subsubsection{Hand-over Authentication Schemes}

In this subsection, we will talk about the hand-over authentication schemes, because it is a very important topic when it comes to the security of mobile applications. In the future, 5G and 6G-enabled networks will contain such applications, therefore, we would like to discuss the existing literature associated with secure handover authentication of these applications. For this, Xue et al. \cite{xue2019secure} proposed a satellite-based authentication model for mobile IoT applications. In this model, the authors bypassed the network control center (NCC) in the authentication process to avoid delay during the verification and validation process of a user. 
In \cite{wang2019sdn}, the authors proposed a novel software-defined networking (SDN)-based handover authentication technique for mobile edge computing-enabled cyber-physical systems (CPS). Specifically, the authentication handover module (AHM) was applied in the SDN controller to ensure key distribution and matching on the client side. The objective of this model was to minimize computation complexities in terms of authentication and validation. Diro et al. \cite{diro2020lightweight} proposed a lightweight authentication scheme for subscriber communication of IoT applications. In this model, the authors used a symmetric-key payload encryption algorithm to ensure the authentication of legitimate users at the end side with the least computation cost. Moreover, the authors checked the proposed model with Transport Layer Security (TLS) for comparative analysis and claimed that their scheme is better than this in terms of computational complexity. 
In \cite{arfaoui2019context}, the author proposed a context-aware lightweight authentication and key agreement scheme for wireless applications. This scheme established authentication between interconnected devices in an employed network. Following this topic, the authors presented a comprehensive survey on this topic in references \cite{ferdowsi2018deep, yang2019faster}.

\subsection{Secure Data Storage}

In Metaverse, stable and reliable operation of the network can improve the productivity of any application. But the convergence of technologies with a shift in the traditional Internet paradigm arises several challenges, and secure data storage is one of them, which needs to be addressed for stability. In Metaverse, the among of data for storage will be massive, and the conventional methods of secure data storage will be not in a position to manage this \cite{sun2022big}. Given that, it is pertinent to develop new big data storage techniques for Metaverse by following the ultra-fast communication paradigm of 5G and 6G-enabled technologies. Sensitive data, such as user, password, biometric, facial recognition data, and behavioral data are susceptible to internal external threats when it comes to the storage of data \cite{zhang2023multiserver}. Despite that, some online transaction information and data of customers and business stakeholders will be stored and accessed dynamically in the Metaverse. For this, some server-based security techniques such as physical security and virtual security approaches have been in the past, as discussed by \cite{adil2023covid}. But still, it is challenging to extend and implement such application-oriented security frameworks in the Metaverse.\\
To explore this topic one step ahead, Ersoy et al. \cite{ersoy2022blockchain} discussed the security storage issues in a decentralized environment by targeting the MetaRepo platform, where users store digital assets such as cryptocurrency, clothes, shoes, avatars, and tickets, etc, in the Metaverse. Furthermore, the authors highlighted the anxiety factors in relation with the security of their own possession's stolen data followed by the legal or illegal transferred data to another application in the Metaverse. Bouachir et al. \cite{bouachir2022ai} present a detailed survey regarding AI-enabled decentralized data privacy techniques in coordination with blockchain technology. In this paper, the authors underlined that it is still under observation how the security challenges will be resolved in the future Metaverse, when it comes to the integration of different applications followed by the real-world transformation to a virtual world. Even though this was a good paper for an overall literature review, but the authors did not talk about the open security challenges followed by their solutions. Therefore, we want to examine the possible open research questions in this article by following the future AI-enabled solutions.     

\subsection{Augmented Reality security in Metaverse }

In the taxonomy of Metaverse, we talked about AR technologies and noted that this technology will include a complex set of input and output entities such as sensors, cameras, displays, earpieces, microphones, etc., followed by GPS to align real-world and virtual-world objects. Furthermore, we observed that this technology will play a key role in the future 5G and 6G-enabled Metaverse. Now, it's time to evaluate the present literature of this technology in the context of security, and check how effective this could in the future. Secondly, we have to identify the limitations of current studies to set the foundation for open research questions.  \\
In the \cite{wang2019blockchain}, the authors talked about immersive attacks in the context of perceptual manipulations. Furthermore, they explained that these attacks are launched through human physical and psychological errors instead of hardware or software vulnerabilities. 
Mhaidli et al. \cite{mhaidli2021identifying} extended this discussion in the context of AR and underlined the methods, tools, and testers that had been used to alter the human wearable sensors and disturb their physical posture to intercept the legal operation of an application. In \cite{gamage2021predictable}, the authors exemplified these attacks by discussing an AR-assisted walk application, where a legitimate user was misguided through such an intervention. It was practically noted that the intercept users were found in the wrong direction.  Likewise, references \cite{kohli2010redirected} and \cite{ohagan2021safety} explored these vulnerability threats in human action predictions followed by the touch screen applications. They investigated that these are the most severe threats to AR, and its associated technologies. Therefore, we believed that these problems should be tackled in the Metaverse by following the Human-Computer Interaction paradigm.

\subsection{Extend Reality security in Metaverse}

The utilization of Extended Reality (XR) in Metaverse is rapidly increasing with remarkable contributions. Given that, this technology became an everyday consumer demand for virtual and real-world object evaluation in Metaverse. Undoubtedly, XR has the potential to help and improve the accessibility of objects in the future Metaverse. But it also opens the door for new risks and attacks to the Metaverse paradigm, where an attacker can easily target individual clients followed by one application or multi-applications \cite{alcaniz2022roadmapping}. In traditional applications, security is a well-established field, but in Metaverse and particularly XR, this field still needs a lot of attention and groundwork, because Human-Computer Interaction (HCI) is evaluated in the recent decade \cite{evans2022white}. This topic is further explored by Tseng et al. \cite{tseng2022dark}, and highlighted the possible risk and threats associated with XR devices in the context of “Virtual-Physical Perceptual Manipulations” (VPPMs). Furthermore, the authors used different attack tactics to hijack the legitimate operation of client wearable devices that uses XR and AR.  In \cite{ohagan2023privacy} the authors underlined that the XR device's operation is susceptible to nearby people, objects, and devices. Furthermore, they performed different experiments to verify their claim. The results, they obtained during simulations, illustrate that this technology is vulnerable to standby objects. Despite, the valuable contributions of this technology in 5G and 6G-enabled Metaverse, we are worried about the security aspects of this technology, and have several open research questions for the involved stakeholders.

\subsection{Virtual Reality security in Metaverse}

From the literature, we noted that Virtual reality (VR) with other associated technologies such as AR, XR, and MR has become a major target of investment for major industry stakeholders, interested in the emerging technology paradigm known as the “metaverse” \cite{lim2022mine}.
Furthermore, we noted that this environment uses special devices to enable the users to interact with the virtual world, and performed different tasks in the digital world. Despite the fact that Metaverse has special promises to bring many things to the doorstep of users. But at the same time, during real-world and virtual-world interactions, the live streaming of users, data is vulnerable to many external attacks. Given that, it has been reported by \cite{duzgun2022sok} that many users' credentials and sensitive data have been captured for malicious activities during transmission in the form of eavesdropping and man-in-the-middle attacks. In \cite{odeleye2022virtually}, the authors suggested that data stored about different users' credentials and activities at the network cloud are vulnerable to data breaches. Furthermore, this topic is explored by researchers in \cite{tricomi2022you}, and illustrated that how adversaries can make fool innocent users, who are playing virtual games to reveal their personal and confidential information. With this, they acknowledge that these attacks are very difficult to handle with the current security measures. Therefore, we would like to explore this area more in the open security challenges section.

\subsection{Summary of Discussion}

In this section, we have discussed different security threats with their counteraction schemes. Given that, we considered the literature of computer networks, WSN, UAVs, ad hoc networks, IoTs, and IIoT, AR, VR, MR, XR, and DT etc., by keeping in view the expected transformation of these technologies into Metaverse. Moreover, we have highlighted the limitations of the adopted security techniques by taking into account different communication and operational factors. To explore, initially, we reviewed the existing attacks with their mitigation techniques to set a concrete foundation for open security challenges with future research directions. With the help of this, we acknowledged the readers the current progress of adopted techniques with their limitations. Given that, we examined the RFID-based authentication techniques to familiarize the involved stakeholders with different frameworks that had been used to ensure the security of existing and future  networks. Despite that, we scrutinized two-factor, three-factor, and multi-factor authentication schemes to admit their contribution and limitations by following the 5G and 6G-enabled Metaverse communication and interconnectivity paradigm. Besides, we talked about some additional security techniques that are dealing with repudiation, non-repudiation attacks followed by interoperability and key matching, etc. With this, we discussed several attacks use cases for AR, XR, MR, VR, and DT technologies in Metaverse. 
The basic notation behind this discussion was to bring new readers, students, researchers, and industry stakeholders with up-to-date progress, and set the door for open research challenges that have been discussed in the subsequent section. 

\section{Open Security Challenges}

\begin{figure*}[ht!]
	\centering
	\includegraphics[width=.85\linewidth, height = 10 cm]{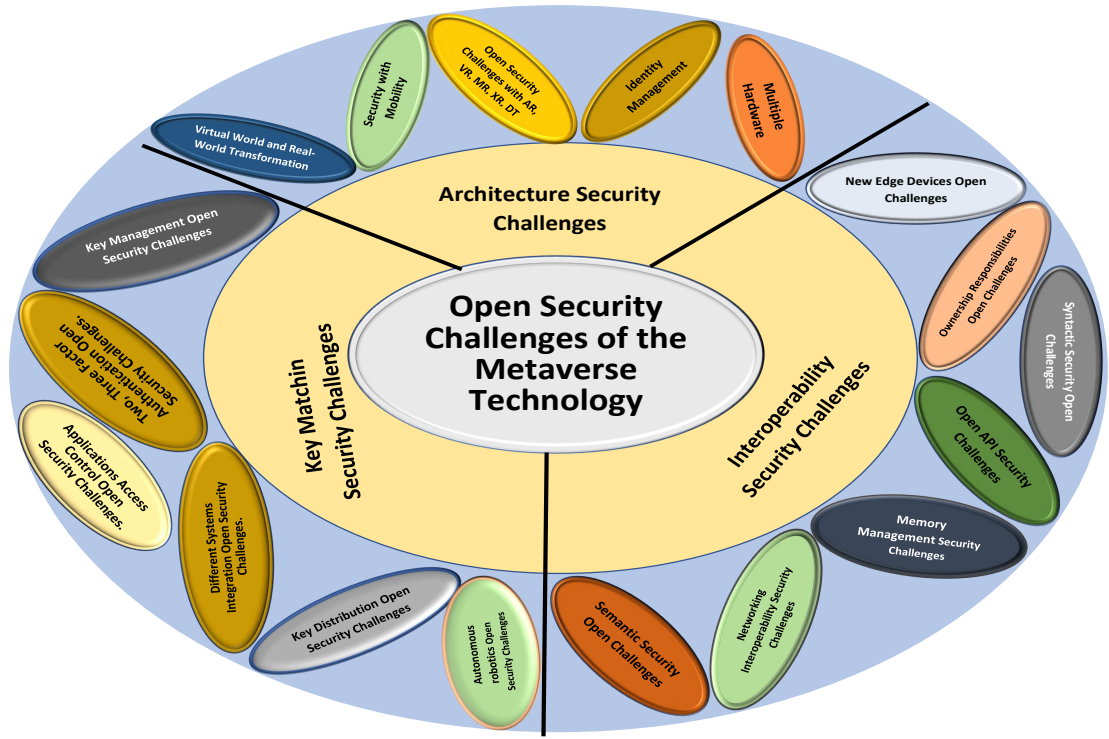}
	\caption{Metaverse technology open security challenges visual representation }
	\label{fig : 8}
\end{figure*}

In this part of the paper, we will explore the open security challenges that still require the attention of the entire community to address and overcome. According to the literature, 5G and 6G-enabled Metaverse technology applications will face a lot of security challenges in the near future, when it comes to the integration of different technologies. Some of them are discussed and summarized in the upcoming subsections. Figure \ref{fig : 8} presents a visual representation of those open security challenges.

\subsection{Open Security Challenges with Heterogenous Network Architectures}

In the future, 5G and 6G-enabled Metaverse technology paradigm is expected to interconnect a variety of applications such as the Internet of Everything (IoE), Internet of Medical Things (IoMT), Internet of Vehicles (IoV), IoT and WSN, etc, under one network shallow. Following this expectation, we have noted in the literature associated with this emerging technology that the existing authentications and data preservation schemes still leave a lot of open holes for an attacker to penetrate in an operational network, when it comes to the different network architectures. Therefore, in this segment, we would like to focus on and highlight the open security challenges connected with the heterogenous network architectures of the Metaverse technology. 

\subsubsection{ Open Security Challenges with Identity Management}

In the future, 5G and 6G-enabled Metaverse technology applications' identity management of devices/clients would be a challenging issue for the research and industry stakeholders, because we have noted in the literature identification of legitimate devices/clients/users is a hectic task. Especially, when it comes up to heterogeneous networks, where billions and millions of sensor devices are connected. It is a common question that how the identity management issue should be handled in the Metaverse technology applications on the client side, because sensor devices had limited resources, which need efficient utilization. To fulfill the cost-effective identity management requirements in these resource-limited network applications, existing authentication solutions would not be able to take care of the computation capability, memory utilization, and onboard power of end-side devices. Therefore, we would like to bring the attention of involved stakeholders toward this serious problem.

\subsubsection{ Open Security Challenges with Mobile Metaverse Applications }

In this segment, we highlight the unsolved security challenges that arose during the mobility of sensor devices in Metaverse technology applications. In mobile Metaverse applications, secure hop-counting communication such as device-to-device, machine-to-machine, or sensor-to-sensor is a challenging task, because the interconnected devices change their locations constantly, and the authentication with the adjacent devices is the utmost requirement for them to transmit their data in the network. Presently, existing authentication schemes are complex in terms of identity verification, matching, and validation, especially, when it comes to the hop count authentication. Therefore, they consume too many resources during the authentication phase. As a result, the computational cost increases, and the network's lifetime decreases. Therefore, we suggest the research community follow up on this problem by designing lightweight authentication schemes that could be useful in mobile Metaverse applications in terms of hop count authentication and handover.

\subsubsection{Open Security Challenges with Multiple Hardware in Metaverse Applications }

Future 5G and 6G-enabled Metaverse applications would be constituted from billions and millions of interconnected devices, as mentioned in the preceding sections. These devices will have different hardware and vendors, which will create authentication problems because the existing hardware-based authentication schemes use default passwords or built-in key-matching techniques. With these techniques, the scalability and interoperability of future Metaverse applications can not be handled properly. Therefore, we suggest the research community work on this open security challenge and devise reliable authentication schemes for this emerging technology.

\subsection{Traditional Networks Attacks and Metaverse Attacks Compatibility Challenges}

In the literature, we discussed different attacks with their counteraction schemes. But we noticed that traditional networks have been targeted through complex attacks, while new technologies such as AR, VR, MR, and XR are targeted via simple, but hidden attacks and most of them have a link with human factors or visible communication aspects. \textbf{Can} we devise new software or hardware to minimize these error rates. Secondly, \textbf{Is} it possible to design a collective security framework for these technologies (new and old) in Metaverse? If yes, \textbf{How} the authentication of different users should be maintained, while using the same device. For how long the authentication parameters of an individual user should be stored in memory, especially, when it comes to the dynamic change of users? \textbf{Does} a large number of authentication parameters in a severs will have an adverse impact on the network? If yes, \textbf{How} this problem should be tackled in the 5G and 6G communication paradigm?

\subsection{Open Security Challenges with Virtual World and Real World Transformation}

In Metaverse, a smooth and seamless transition between virtual and real-world objects is a very important aspect. When new users, especially, those who do not have sufficient knowledge of this technology, \textbf{How} they can be secured from unnecessary and vulnerable advertisements. \textbf{Is} possible to create a platform, where users can differentiate between legal and illegal sites and advertisements. Moreover, \textbf{How} the authentication process should be made simple and easy for a user, when he/she is propagation between virtual worlds. Moreover, the new virtual world transition demand authentication parameters of a user. Then, \textbf{how} this process should be made transparent, we mean, \textbf{How} it should be assured to them that his transition to the new virtual platform is safe. As there is much possibility of the alike fake virtual world. it is essential to respond to these questions and design a secure real and virtual world platform in the Metaverse.

\subsubsection{Open Security Challenges with APIs, AR, VR, MR and XR}

Application programming interface (API) plays an essential role in the interoperability of any technology application because it provides an interface for services intending to expose data to high-level languages. Existing APIs, such as  RESTful/REST, IFTTT, Open IoT, and LinkSmart11, etc., are designed and secured in the context of different applications requirement [180]. When it comes to the heterogeneity of Metaverse, where numerous special devices of VR, AR, MR, and XR will work together. Then \textbf{How} their security should be ensured? \textbf{What} tool and testers should be used to guarantee the data integrity, confidentiality, and availability in an operational network? Despite that \textbf{What} kind of preventive measure should be taken to ensure the security of devices with the portability of applications such as IoT, IoE, IoV, and many more in coordination with AR, VR, MR and XR devices?   

Therefore, we suggest the research community target the security issues associated with the interoperability of Metaverse technology APIs.

\subsection{Open Security Challenges with Interoperability of Metaverse Applications}

In this section, we talk about the open security challenges associated with the interoperability of Metaverse technology applications. Future 5G and 6G-enabled Metaverse technology will contain numerous IoT, WSN, UAVs, IoV, and IoE applications, which will have different security requirements, as discussed in the literature. To explore these open security challenges, we classify them into two sub-categories: semantic and syntactic interoperability, respectively.     

\subsubsection{Open Security Challenges with Semantic interoperability}

The integration of numerous applications under one 5G or 6G-enabled network paradigm of Metaverse technology will raise semantic interoperability challenges. Redressal of these problems with existing interoperable technologies will raise several security challenges for the research community because each application has different service requirements. To this end, we intend to highlight the open security challenges that can play a pivotal role in this emerging technology operation. Specifically, the first and foremost challenging task would be to enable the secure distribute-ability, accessibility, and operation-ability of eXtensible Markup Language (XML) for sensor devices by taking into account their limited resources and application requirement. Secure database management is an open security challenge because this emerging technology is an ingredient of several networks/applications, which have different operational requirements.  Likewise, the lack of a standardized semantic security platform for such a heterogeneous network paradigm is another challenging task for the research and industry stakeholders.

\subsubsection{Open Security Challenges with Syntactic Interoperability}

In [181], the authors discussed the operational requirements of syntactic interoperability in the context of security concerns by taking into heterogeneous networks. After evaluation of this article in the coordination of existing authentication and data preservation schemes, we have concluded that Metaverse technology will offer several security challenges when it comes to the interoperability of different applications under one network paradigm. Herein, we would like to underscore some of the open security challenges, which need considerable attention from the research community and involved stakeholders. When the sender and receiver have different coding and encoding schemes, how secure should the transmission be among paired devices, especially when it comes to resource-limited devices such as sensors, IoT, IoE, etc? Besides, is it possible to enable device-to-device authentication with hop count communication among different applications under one network paradigm?

\subsection{Open Security Challenges with 5G-and 6G-enabled-Mateverse and AR}

In the literature, we noted that AR technology equipment, gadgets, and devices should be targeted through human low-security pools in different applications. With this, We also witnessed that different matching techniques have been used to ensure the security of these devices during dynamic connectivity. However, we are concerned, \textbf{How} these devices and applications will respond, when it comes to the multi-application operation with 5G and 6G enabled-ultra-fast communication of Metaverse. Despite that, \textbf{Is} it possible for a user to authenticate with the same password/security parameters followed by
utilized gadgets or devices in different Metaverse applications? If yes, \textbf{How} much secure this framework would be? If not, then, \textbf{What} are the possible options for the solution to this problem? In addition, \textbf{Is} it possible to merge the existing AR applications with newly adopted applications with proper security protocols in Metaverse. If yes, \textbf{How} this platform should be checked? If not, \textbf{What} we can do for redressal of this.

\subsection{Open Security Challenges with VR technology in 5G and 6G-enabled Metaverse}

In the literature evaluation phase of VR, we noted that this technology is susceptible to various attacks. Some of these may be the result of mistakes made by people, while others may be brought on by the lack of client awareness. To follow up on these threats, we did not notice considerable countermeasures techniques that have the capability to prevent the interception of adversaries in future Metaverse using VR. Given that, we have some questions for the involved and even clients. \textbf{How} transmission of VR-enabled devices should be secured or segregated from the side by devices? \textbf{Is} there any possibility of this? If yes, \textbf{How} it could be done? Does it will have an impact on the performance of VR gadgets or not? Secondly, \textbf{How} the customer's awareness problems should be resolved? \textbf{What} are the possible options that could be utilized to acknowledge the proper use of VR devices, especially, when it comes to a massive number of users? Despite that, \textbf{Can} we launch an easily accessible and understandable security coursework for the users? If yes, \textbf{How} uneducated people will be dealt with? If not, then \textbf{How} we can secure the foolproof security of VR in Metaverse? 
5G and 6G ultra-fast communication are proven some external attacks, when it comes to the integration of VR devices with existing networks (using old communication entities). Then, \textbf{What} are the possible techniques that can be adopted to ensure the compatibility of VR devices with already established networks using 5G and 6G communication paradigms? 

\subsection{Open Security Challenges to Online Gaming Industry}

In the future Metaverse, the gaming industry is expecting considerable growth in the context of the economy. However, they are needed to ensure the reliable operation of their applications, as discussed in the gaming industry section. Despite that, another challenging task for these stakeholders would be the security of the consumers market individuals. Because most of the users will be connected dynamically followed by static and mobile network paradigms. When this connectivity is dynamic with the mobile network, then, \textbf{How} the authentication of an individual should be ensured by keeping in view the efficient utilization of network resources. Given that, this authentication should be centralized or decentralized. Let's assume, if it is a centralized authentication framework, then, \textbf{How} resource-limited networks such as IoT and WSNs should be dealt with in the context of resource management and traffic congestion and contention. Conversely, if it is a decentralized framework, then, \textbf{How} the authentication parameters a legal user will propagate in the network? \textbf{How} the memory management issues will be tackled? These are the most important questions, which need a response from the concerned stakeholders.

\subsection{Open Security Challenges with Network Virtualization}

In the 5G and 6G-enable Metaverse, virtualization is a key concept regarding the physical world and virtual world integration. Given that, this virtualization should be ensured through numerous software and hardware combination and compatibility. When a particular software is compromised through malware, then, \textbf{How} the propagation of this malware should be restricted to adjacent software. During virtualization, \textbf{Is} it possible to design a layer-wise security framework? If yes, then, \textbf{How} constraint-oriented networks should be dealt with. We mean, \textbf{There} should be different security approaches for these networks. If not, then, \textbf{How} this problem should be managed?
Despite that, \textbf{How} the access control policies should be devised for this paradigm? Generally, an admin can manipulate such vulnerability threats. But in the case of Metaverse virtualization, \textbf{Who} will take this responsibility? Secondly, it should be application-specific or Metaverse. Despite that, if it is application-oriented, then, \textbf{What} consequences it will have on the network performance? if it is Metaverse base, then, \textbf{How} the different application's security parameters should be managed under one umbrella?

\subsection{Human Centric and Content Centric Metaverse}

In the future Metaverse, a large number of user-generated contents (UGC) are envisioned to be constructed, ordered, and delivered across several sub-metaverse technologies. From the current literature, we can see that IP-based content transmissions are facing many imperative challenges to secure UGC data during transmission in large-scale networks. 
To compensate for these challenges, the literature presents several counteraction schemes, but it is still not clear, \textbf{who} will be the trusted party for all interconnected applications during content dissemination when it comes to the integration of large-scale different networks. Secondly, \textbf{Is} it possible for all sub-networks to accept such a paradigm? If not, then, \textbf{What} are possible options that can be useful to gain the trust of all involved stakeholders?

\subsection{Open Security Challenges with 5G-and 6G-enabled Autonomous robotics and Systems in Metaverse  }

In Metaverse, it is expected that 5G and 6G-enabled autonomous robots and systems will be merged into this technology. Despite the fact that this technology will contribute a lot to the economy, but it is worth noting that \textbf{How} the interoperable sensor should be secure from internal and external adversaries. Especially, when it comes to malware propagation, spyware, and trojan. \textbf{Is} it possible to stop the propagation of aforesaid viruses among interconnected sensors? If not, then, \textbf{How}  the adjacent sensors should be made secure from them? If yes, \textbf{What} are the tools and testers that could be useful prevent the spread of these viruses? 
Moreover, \textbf{Is} it possible to manage the operation of different sensors separately? If yes,then, \textbf{How} the synchronization among different sensors should be maintained with reliable operation?

\subsubsection{Open Security Challenges with Networking Interoperability}

Future 5G and 6G-enabled Metaverse technology will allow the interconnection of numerous applications under one network shallow. In the literature, we have noted most of the existing authentication and privacy schemes are application specific. How could such an enormous network paradigm be secure during different applications' integration when it comes to the network layer, session layer, and application layer protocols? Moreover, we know that the evaluation of this technology is still in the developing phase, therefore, it is an open challenge for the research community to build a standardized security framework that could be capable of evaluating the performance of different routing protocols in coordination with security techniques.

\subsection{Open Security Challenges with Key Matching}

Future 5G and 6G-enabled Metaverse Technology applications will use key matching schemes for device-to-device (D-2-D), machine-to-machine (M-2-M), and sensor-to-sensor (S-2-S) authentication under one network paradigm at the client side. Based on the literature evaluation, herein, we would like to highlight the open security challenges that will be aroused in the Metaverse technology. For this, we considered the heterogeneity and scalability of the network followed by their integration with different applications.
In the consequent sections, we will have summarized them in connection with existing literature.  

\subsubsection{Open Security Challenges with Key Distribution}

In the future, 5G and 6G-enabled Metaverse technology applications, key distribution will create a lot of challenges for the research community, because there will be connected a lot of applications under one network shallow.  With this platform, \textbf{how} the authentication keys should be distributed by keeping in view the mobility factor. And \textbf{how} the D-2-D authentication should be enabled among different applications clients at the end side? Moreover, it is also challenging that \textbf{how} public and private keys should be distributed among the interconnected devices in such a heterogeneous network. Likewise, \textbf{how} the default authentication scheme should be dealt with when it comes to the key matching of different applications. These are the most visible challenges associated with the key distribution, verification, validation, and matching of the Metaverse technology.

\subsubsection{Open Security Challenges with Key Management}

From the literature on Metaverse technology, we have noted that this technology should be extended from specific to more general applications in the form of heterogeneous networks. Following this infrastructure, how the public and private keys of network participating entities should be managed because there will be different devices that will have distinct computation capabilities? To exemplify this, let's assume sensors, PCs, laptops, smartphones, etc., \textbf{who} will be responsible for managing such a huge number of authentication keys? In the context of hop count communication, \textbf{how} the resource-limited devices will maintain their pre-store register in the context of neighbor devices keys, if they follow hop count communication? With this, \textbf{how} the communication and computation capabilities should be retained to achieve the desired quality of service requirements in Metaverse technology applications.

\subsubsection{Open Security Challenges with Two and Three-Factor Authentication}

In the present literature, we noticed that most of the two- or three-factor authentication schemes are application-specific. Therefore, it would be a challenging ask for researchers to enable two or three-factor authentication among numerous interconnected applications (devices) under one network platform for the Metaverse technology. In particular, \textbf{how} two- or three-factor authentication should be enabled among different application devices in one network paradigm such as Metaverse technology? \textbf{What} are the parameters that would be adopted during the authentications? \textbf{How} these two- or three-factor authentication should be managed, when it comes to the mobility of devices in an application? \textbf{Would} these two or three-factor authentication schemes be communication friendly in Metaverse technology applications? Following the mobile authentication, handover process, and communication metrics of Metaverse technology, the research community needs to work hard on this challenging problem.

\subsection{Open Security Challenges with Addition of New Edge Devices}

In the future 5G and 6G-enabled Metaverse applications, it is likely accepted that some additional devices such as glasses (uses AR or VR), goggles, joysticks, and headphones, etc., should be used to connect with the edge devices to satisfy the customer's needs in different applications. To exemplify, let's assume the scenario of virtual gaming applications of Metaverse technology. When it comes to connectivity, operational ability, and client participation in these applications, the aforementioned devices would be used as additional constituents to facilitate end users at their doorstep. To explore this, we can see that the highlighted apparatuses have a very important role in the Metaverse technology productivity and user satisfaction. While at the same time, we can also expect that it will offer several open security challenges for the people working in this domain. For example,  \textbf{how} the authentication should be ensured of different external devices with edge devices during first connectivity; \textbf{how} data privacy should be maintained in such a scenario; \textbf{how} should its complication be minimized in the whole Metaverse technology paradigm if a device is compromised? Likewise, is it possible to identify an adversary in such a scenario during forensic investigation? These are open security challenges that need the attention of involved stakeholders if we want a safe, secure, and reliable future Metaverse technology paradigm.

\subsection{Open Security Challenges with Different Systems Integration}

From the literature and future perspective of Metaverse, we can observe that this technology will integrate a large number of applications under one network paradigm that could be visible in terms of the digital transformation of the globe. When it comes to the integration of such a great number of applications, then \textbf{How} the security of these different applications should be managed under one network platform. Given that, \textbf{Is} it possible to ensure the security of such different applications? If yes, then, \textbf{What} should be the authentication framework for them? If not, then \textbf{How} we can integrate different applications in Metaverse? These are the basic questions for the research community and industry stakeholders, which need their responses.

\subsection{Open Challenges with the Security Ownership Responsibilities}

In the evaluated literature, we have not come across a single paper that deals with the ownership of the security of this emerging technology. To explore, when a subnetwork or application is compromised, which network entity should be considered during an investigation or forensic, local network or global network, and \textbf{Who} will take the ownership of security. Despite that, \textbf{What} should be the defined parameters for that network entity in intra-networking and inter-networking? Secondly, \textbf{Is} it possible to manage the security of the whole Metaverse parading through one network entity and it could be considered during forensic verification? If yes, then \textbf{How} the internal security approaches of different applications should be dealt with? If not, then, \textbf{What} are possible options that could be adopted to handle this problem? In the case of a centralized entity, \textbf{What} are possible options exist that could be used to check the feasibility?  Conversely, if it is decentralized, then, \textbf{How} the situation will be dealt with, when a subnetwork or application is compromised through other networks? 
We believe that these are the most demanding questions, which need the concerned stakeholder's response to guarantee reliable forensic investigation of this technology.

\subsection{Open Challenges with Security and Memory Management}

In Metaverse, different applications should be connected under one network umbrella. Therefore, efficient utilization of memory with proper security parameters should be a challenging task, and we have noted this in the literature review phase. Given that, in Metaverse, there will be computer networks and sensor networks such as Wireless Sensor Networks (WSNs) Unmanned Aerial Vehicles (UAVs), Ad Hoc Networks, and the Internet of Things (IoT), etc, that have limited sources. Keeping in view the limitations of these networks, \textbf{Is} it possible to adopt a global security framework for Metaverse? If yes, \textbf{How} should the processing capabilities and memory concerns of the limited resource network will be dealt with? If not, then, \textbf{What} are the possible alternatives that could be useful in the redressal of this problem? Despite that, If lightweight authentication techniques are adopted for resource-limited networks by taking into account their memory storage and processing powers, then, \textbf{How} efficient these techniques could be in computer networks. We mean, these techniques will be capable to counter different attacks that had been exercised on computer networks.        In the case of inability, being a member of community, we should have to think about the better future of this technology.

\subsubsection{Open Security Challenges to the Metaverse Economy Market}

Metaverse will enable different applications stakeholders to present and ensure their client's participation and products under one network platform, in most cases openly in the form of digital asset trading and resource sharing. With this, they will boost up the reward of their productivity in the context of increasing overall gross and net profit. Although this technology offers several benefits to business stakeholders in the digital market, but, at the same time, it offers numerous challenges to the economy of these enterprise market owners. To exemplify these challenges, the undermentioned scenarios should be kept in mind. 

\begin{enumerate}
	\item 
	The first and foremost security challenge is, \textbf{How} the authentication parameters of legitimate clients can be maintained in different applications, especially, those they are openly accessible in Metaverse paradigm.  \textbf{How} to restrict unfair free access to different applications of this technology by keeping in view the ease of access and connectivity policies. \textbf{What} are the security parameters that should be adopted for this technology without losing simplicity during users connectivity? 
	
	\item If an application is compromised, \textbf{How} the economic damage can be controlled? \textbf{Is} it possible to minimize the damage to a specific domain in an application? If yes, \textbf{What} should be these preventive measures? \textbf{Does} it will have an impact on the performance of the network? If not, then, \textbf{How} the trust of enterprise market stakeholders can be gained? Secondly, \textbf{How} it should be demonstrated to involved consumers and business stakeholders that their investment is secure.

\end{enumerate}

\subsection{Open Security Challenges with the Different Applications Access Control}

In Metaverse, the next biggest security challenge is associated with the access control policies of different applications. When unlike applications have different access control policies, then, \textbf{How} the session initiations and storage should be securely controlled, especially, when an admin login through a different application. Despite that, the login cookies of an administrator are saved in an irrelevant server, then, still, a particular application is safe from adversaries. If not, then, \textbf{What} are the possible ways through which this problem can be handled in future Metaverse?  Likewise, \textbf{Is} it possible to manage access control policies with respect to applications and subnetworks. If yes, then, \textbf{How} the easy-to-access formality should be fulfilled. If not, \textbf{What} are possible solutions?

\subsection{Open Security Challenges with Traditional Communication and 5G and 6G Communication Hardware (Physical Layer)}

From the taxonomy of Metaverse and 5G and 6 G-enabled technologies, we can observe that when the existing technologies are merged in Metaverse, then, \textbf{How} the communication, QoS, Interoperability, and security challenges can be dealt with. To explore, presently, we can see that mm-wave bands have been widely used in 5G and 6G communication infrastructure. In Metaverse, when such a communication paradigm is adopted, then \textbf{How} the physical layer security of traditional applications should be handled? \textbf{Is} it possible to make the conventional application's hardware and software compatible with 5G and 6G capable hardware and software? If not, then, \textbf{What} are the possible options that could be useful to make this process with proper security protocols?

\subsection{Open Security Challenges with Traditional Communication and 5G and 6G Communication Hardware (Data Link Layer)}

When it comes to data link security, we noted that in the 5G and 6G communication paradigms, the narrow bandwidth with short pulse transmission waves will minimize the eavesdropping attack ratio. But when it comes to the traditional network integration with 5G and 6G communication technology, then, \textbf{How} the channel coding should be managed. Secondly, let's suppose, we manage the transmission channel with the help of coding, then, \textbf{How} secure these channels should be? Because it is like opening a door for an adversary. \textbf{How} this scenario should be manipulated?
Thirdly, In 5G and 6G communication, the attenuation of THz communication is very low for certain materials, while it is high for traditional 2G, 3G, and 4G communication infrastructure. During merging, \textbf{How} this problem will be tackled? If it is not managed properly, then, it creates an environment of jamming or eavesdropping attack. Given that, \textbf{How} it should be ensured that deployed Metaverse network communication paradigm is secure and safe?

\subsection{Summary of Discussion}

In this section, we talked about the different security challenges that still require the concerned stakeholder's response for a better future of Metaverse. Based on the existing literature limitations and Metaverse requirements, we have taken into account several aspects of existing technologies followed by emerging technologies, while managing this segment. Because it was important to synchronize the requirement section and existing literature sections with open security challenges, as Metaverse will integrate many technologies that range from old to emerging. Furthermore, we considered the operation, additional device participation, communication, and interconnectivity scenarios from the client's side to the network cloud followed by the remote destination. Given that, the questions raised in each subsection have a direct link with 5G and 6G-enabled Metaverse. Therefore, we strongly believe that the redressal of these questions can solve the security problems of future Metaverse up to a significant level.

\section{Future Research Directions}

In this section, we highlight the possible research directions that could be useful in the redressal of the aforementioned open security challenges.

\subsection{ Future Directions for Authentication of Individual Devices}

In previous sections, we discussed that the authentication of participating entities in Metaverse technology applications would be a challenging task, due to the heterogeneity, mobility, scalability, and handover processes. Also, we showed that default password techniques could not be an effective way to ensure cost-effective authentication in the future 5G and 6G-enabled Metaverse technology applications, because of the aforementioned factors. Despite this, some devices such as smartphones use SIM cards for authentication in the network. In the future Metaverse technology applications, it would not be possible to ensure the authentication of billions and millions of devices through SIM cards and default passwords, therefore, it's time to start the utilization of E-SIM cards.

\subsection{Future Research Directions for Cross-Domain Integration in Metaverse}

To compensate for the cross-domain trust issues in Metaverse, blockchain could be used as a promising technology, because it has the capability to build a trust-free economy ecosystem for this technology. Blockchain will allow distinct applications or sub-metaverses to deploy their services such as block structures, transaction formats, and consensus protocols to meet the security and network interconnectivity requirements in the context of interoperability. Given that, an efficient cross-domain authentication framework will ensure the legitimacy of participating consumers followed by digital asset-related activities in different sub-metaverses. Although currently hash-locking, side-chain, relay chain, and digital asset transfer security techniques had demonstrated remarkable results in blockchain technologies, but, it has much potential and can do better than the present in the future Metaverse. Therefore, we believe that the adaptation and execution of this technology should be further investigated in the cross-domain connectivity of Metaverse by targeting the application-specific requirements and network architecture. 

\subsection{Future Research Directions in Gaming Industry}

In the future gaming industry, the players or clients will be monitored or facilitated through wearable devices such as goggles, smartwatches, sensor-embedded shoes, and cloths that will be resource-constrained in terms of computation and communication capabilities, as discussed in the literature. Once this technology is merged into Metaverse, which always demands resource-full devices that has high computation and communication power to ensure delay-sensitive connectivity and transmission. Therefore, data privacy with the authentication of dynamic players and devices is a challenging task for the involved stakeholders. But, we trust that user-generated content (UGC) that uses content management nodes in a decentralized environment, and CCN could be used as a promising technology effectively addressed the security concerns of this technology.

\subsection{Future Research Directions for Scalability of Metaverse Technology}

In the open security challenges, we highlighted the scalability problems associated with the security concerns of Metaverse technology applications. Although, this problem is associated with the long-term planning of the Metaverse technology paradigm, it's a good time to suggest possible research directions that could be adopted for redressal of them. Machine learning (ML), Deep Learning (DL), and Reinforcement Learning (RL) algorithms should be used as an effective weapon against new, previous, and unknown threats because these algorithms are capable of identifying them cost-effectively in heterogeneous networks. Moreover, Transfer Learning (TL) algorithms could be used to address the security problems associated with Metaverse technology, because they are capable of defense distillation (a kind of defense strategy) to ensure the security robustness of this emerging technology. Therefore, we suggest the involved communities focus on the utilization of ML, DL, RL, and TL algorithms in future work to detect and prevent complex security threats of heterogeneous Metaverse networks.

\subsection{Future Research Directions Human Centric and Content Centric Security Challenges}

In the recent past, Content-centric networking (CCN) emerged as an alternative technology that shifted the traditional Internet architecture. In CCN contents are managed and routed with respect to their naming addresses, however, these content are manipulated through host-oriented routing paradigms and IP-based routing protocols in the traditional Internet architecture. To explore, in CCN-based paradigms, the UGC consumers would be enabled to send the desired UGC request to any CCN node network that has close matches with UGC. In addition, the CCN nodes have embodied a security model which explicitly checks and verifies the security of individual content instead of securing the “pipe”. Although, this technology is still in its emerging phase, but keeping in view its potential,  we believe that it can resolve the security concerns associated with content dissemination and handling in the future Metaverse. Despite that, we also noted that CCN can bring some problems such as content poisoning in very large-scale networks, but it can be handled through proper research collaboration between industry and academia. Despite that, the metaverse is human-centric, whereas an individual such as users/avatars’ personalized privacy techniques should be in collaboration with CCN to ensure foolproof security of the future Metaverse.

\subsection{Future Research Directions for Security Standardization Framework}

From the present literature, it is clearly visible that there is no standardized security platform for emerging technologies and particularly Metaverse technology applications. Furthermore, it would be even more challenging to design an international security paradigm for the future 5G and 6G-enabled Metaverse technology, because the commercial deployment of the 6G network is expected to start in 2028 in Korea, and then expanded to the globe by 2030 \cite{far2021laptas, amintoosi2021tama}. Therefore, we suggest the research community and industry stakeholders work on the security aspects of this progressive technology and take an extensive part in the development and deployment of 6G technology. With the adoption of such a global security platform, numerous security challenges of future Metaverse technology can be resolved by taking into account real network scenarios.        

\subsection{Future Research Directions for Forensic Investigation Challenges}

In the future 5G and 6G-enabled Metaverse, it is likely expected that the analogy of social norms should be adopted the same as of present technology. But Metaverse will allow additional accessibility and interaction with several other technologies, whereas content creation, social activities, and business advertisement followed by virtual economy should be very easy for every individual using this technology. However, we noticed in the literature and open research challenges, \textbf{How} the culprit or an adversary should be identified in forensic investigation, \textbf{Who} will take responsibility for a central point of interest in this investigation. Then, we were out of concrete responses/answers. To compensate for these issues and ensure reachability to the culprit, we believe that the world has to review the existing rule and regulations. Furthermore, they have to make them more strict, and devise some punishments for application or sub-metaverse stakeholders as well, because this will be a step forward procedure for them to verify the legitimacy of their consumers. Despite that, blockchain should be used as an additional constitute, and the involved enterprise market stakeholders must be assured to use this technology in coordination with their local technology. With the help of this, the authentication parameters of participating customers should be saved in a decentralized environment cost-effectively, and could be used during a forensic investigation, if needed.

\subsection{Future Directions associated with Reconfigurable Intelligent Surface}

In the open security challenges, we discussed that future Metaverse technology applications are facing many security problems, because of the wireless communication, mobility of devices, and random interconnectivity among the network participating devices. To address these challenges in future 5G and 6G-enabled Metaverse technology applications, the role of the reconfigurable intelligent surface could not be ignored, because of its flexible communication capabilities. Therefore, we suggest the researchers and industry stakeholders work on the security aspects of future Metaverse technology applications to ensure the effective utilization of reconfigurable surfaces in the redressal of these problems. Especially, this technology will be very helpful against man-in-the-middle attacks and jamming attacks. It is also believed that the effective utilization of this technology could be assumed to be an effective weapon to improve the productivity of any application in the Metaverse technology paradigm.

\subsection{Future Research Directions in AR, VR, and XR }

When it comes to the utilization and operation of AR, VR, MR, and XR, several open security have been noted and raised in different sections of this paper. However, it is more challenging to demonstrate an accurate research direction that could be useful in the redressal of underlined open research questions. Given that, some of the security problems such as authentication of devices/gadgets of AR, VR, MR, and XR can be addressed with the utilization of Transfer learning algorithms. Likewise, deep learning algorithms and reinforcement could be used as an alternative paradigm. With this, we suggest that a blockchain paradigm with supervised and unsupervised machine learning should be as used a third candidate to manage the authentication and data preservation problems. Despite these all, we strongly recommend the involved stakeholders to design some training and security awareness campaigns for all customers. Because digital transformation allows attackers to target clients through Phishing attacks (social engineering attacks). With the adaptation of such a basic knowledge platform, most of these attacks can be easily avoided by customers. In short, we trust that aforesaid research directions can be very helpful in the solutions of considered technologies security problems.        

\subsection{Future Directions for Key Management in Distributed Environments}

In the future 5G and 6G-enabled Metaverse technology applications, the key distribution would be a challenging task, as mentioned above in the open security challenges. For the redressal of this problem, blockchain technology should be assumed as the best candidate, because it will allow the authentication of network devices in a distributed environment. However, key distribution with respect to device location is still an open issue. Therefore, we suggest the research community properly utilize ML-enabled algorithms for this task by taking into account the blockchain technology infrastructure. With the help of these algorithms, they must train their models to shift the keys of interconnected devices with respect to device mobility/location. Besides, the virtualization of keys should be considered another solution, because it will ensure the effective utilization of network resources in a distributed environment. In addition, virtualization should be very helpful in improving the overall productivity of any application connected to Metaverse technology.

\subsection{Future Research Directions Associated with Malware Propagation in Autonomous robots and systems}

To compensate for the virus propagation issues in 5G and 6G-enabled Metaverse applications such as autonomous robotics and systems. We suggest the research community to design a firmware build-in firewall for machine and robot-implanted sensors, that have the capability to analyze the incoming and outgoing data. Despite that, we want to acknowledge that internal and external intrusion detection systems (IDS) and intrusion detection and prevention systems (IDPS) could be also very useful in the redressal of this problem. Before the implementation of this technology, it is needed to investigate and testify the new traffic analysis techniques for several possible attacks. Moreover, transfer learning (TL) and deep learning (DL) algorithms could be very productive too, because it has the capability to predicate and detect malicious traffic.

\subsection{Future Research Directions with Wireless brain-computer}

To address application-oriented or sub-metaverses security challenges in the future 5G and 6G-enabled Metaverse technology. We suggest the research community and industry stakeholders to explore Wireless Brain-Computer Interaction (BCI) technology because this technology has considerable potential to address the authentication and data privacy issues in large-scale Metaverse. Although BCI is not new technology and is used in the healthcare domain, but, it has not been explored in the recent past. To elaborate, BCI technology works like the visual cortex in the form of artificial limbs, which uses a four-step process such as signal acquisition followed by feature extraction, translation, and report response. Following the aforesaid four parameters, we believe that these can be useful in the authentication of machine-embedded sensors if utilized in the industry. Despite that, it can be also useful to detect malicious data at the client side and centralized location, if adopted effectively.

\subsection{Future Research Directions for Addition of New Devices}

To address the security concerns that arose with the interconnectivity, operation-ability, and customer participation through external devices in the Metaverse technology applications. We suggest the research and industry stakeholders work together and design such a hardware and software infrastructure for the edge devices that could be capable of not allowing the further process of external devices' data in the network except the defined commands/parameters. Although this is extremely hard to manage such a scenario in the real network, we believe that ML, DL, and RL algorithms could be utilized as supplementary candidates to analyze the upcoming traffic of external devices for the aforementioned task handling. With the collaboration of these two ingredients, the anomalies should be removed from the data, and the security parameters of the network should be maintained to attain the trust of involved stakeholders. 

\subsection{Summary of Discussion}

In this subsection, we talked about different research directions by taking into account the open security challenges of the Metaverse. Even though, some of the solutions need the collective efforts of different stakeholders, but still, we believe that this is possible. Despite that, we also highlighted some solutions that have direct a link with the clients. But we want to acknowledge that to ensure the foolproof security of Metaverse, we need to address all the possible flaws.   

\section{Conclusion}

In this paper, we have presented a comprehensive survey regarding the future 5G and 6G-enabled Metaverse technology applications' taxonomy and their security concerns. Firstly, we covered the literature associated with the taxonomy of Metaverse and 5G and 6G-enabled technologies followed by their different applications to familiarize the readers with the importance and potential of this emerging technology. Then, we narrowed down our discussion to different security threats to the Metaverse technology and their counteraction schemes to accentuate their benefits and drawbacks. Afterward, we set a synchronized preface to underline the requirements of this emerging technology by taking into account the limitations of the present literature. Based on this, we highlighted the open security challenges that still need the concerned stakeholder's attention to ensure the robustness of this technology. Finally, we spotlighted the possible research directions that could be useful in the redressal of underlined open security challenges. Keeping in view the potential of this emerging technology and security challenges, we believe that this paper could be considered as a complete package for the students, industry stakeholders, and researchers, and guarantee the security of this progressive technology in the future.

\bibliographystyle{abbrv}
\bibliography{refs}
\end{document}